\documentclass[10pt,bibliography=totoc]{article}
\usepackage{soul}
\usepackage[utf8]{inputenc}
\usepackage[english]{babel}
\usepackage{graphicx,amsmath,amsfonts,amssymb,dcolumn,amsthm,slashed,bbm,xspace,pgffor,mathbbol,latexsym,cite,braket}
\usepackage{microtype}
\usepackage{lmodern}
\usepackage[normalem]{ulem}
\usepackage{xcolor}
\usepackage[colorlinks=true,citecolor=blue,urlcolor=blue,linkcolor=blue]{hyperref}
\usepackage[hang]{footmisc} 
\usepackage[affil-it]{authblk}

\usepackage{multicol}

\numberwithin{equation}{section}

\begin{document}

\title{\textbf{Spontaneous symmetry breaking in a non-Abelian topological
gauge theory }}


\author{Octavio C. Junqueira$^{a,b}$\thanks{\href{mailto:octaviojunqueira@gmail.com}{octaviojunqueira@gmail.com}},~
Rodrigo F. Sobreiro$^c$\thanks{\href{mailto:rodrigo\_sobreiro@id.uff.br}{rodrigo\_sobreiro@id.uff.br}},~ Nelson R. F. Braga$^a$\thanks{\href{mailto:nrfbraga@gmail.com}{nrfbraga@gmail.com}}}
\affil{\footnotesize $^{a}$ UFRJ --- Universidade Federal do Rio de Janeiro, Instituto de Física,\\
CEP 21941-909, Rio de Janeiro, Brasil\\
$^{b}$ UFABC --- Centro de Matem\'atica, Universidade Federal do ABC,\\ 09210-580, Santo Andr\'e, Brasil\\
 $^{c}$ UFF --- Universidade Federal Fluminense, Instituto de F\'isica,\\ Av.~Litoranea s/n, 24210-346, Niter\'oi, RJ, Brasil}

\date{}
\maketitle

\begin{abstract}
We study the spontaneous symmetry breaking mechanism in a non-Abelian topological gauge field theory, built from the twisted $\mathcal{N} = 2$ super-Yang-Mills theory in the presence of a Fujikawa-type potential. Specifically, by employing Fujikawa's Becchi-Rouet-Stora-Tyutin method, local degrees of freedom are released from the introduction of a potential in the trivial sector of equivariant cohomology. Such a potential displays a nontrivial vacuum solution, which induces the spontaneous symmetry breaking of the gauge symmetry together with the original fermionic scalar supersymmetry of the topological action. In this case, not only massive vector bosons emerge, but also fermionic fields with massive poles. This result shows that the introduction of a topological phase in non-Abelian gauge theories could provide a mechanism of mass generation for fermions with their masses correlated to the mass of Higgs gauge bosons ($m_B$). For the SSB in the topological case, three different vacuum directions are required. Otherwise, the supersymmetry could not be broken, and mass generation for fermions will not occur. Starting with a theory with symmetry $G = SU(N)$, to obtain a gauge theory at the end of the process, we must have $N \geq 3$. We study a maximal symmetry breaking of the type $SU(3) \rightarrow U(1) \times U(1)$, and obtain their fermionic propagators with mass poles $m^2_F = m^2_B = v^2$ after SSB, being $v^2 $ the energy scale introduced by the Fujikawa-type potential.

\end{abstract}

\section{Introduction}

Based on the self-dual Yang-Mills equations introduced by A. Belavin, A. Polyakov, A. Schwartz, and Y. Tyupkin in the study of instantons \cite{Belavin:1975fg},  S. K. Donaldson discovered topological structures of polynomial invariants for smooth 4-manifolds \cite{DONALDSON1990257, Donaldson:1983wm, donaldson1987}. Inspired by M. Atiyah's conjecture \cite{Atiyah:1987ri, Witten:1988ze}, E. Witten found a relativistic field theory, namely, the twisted $\mathcal{N}=2$ super-Yang-Mills theory (SYM) that is capable of reproducing the Donaldson polynomials in the weak coupling limit \cite{Witten:1988ze}. This topological quantum field theory (TQFT) is commonly referred to as the Donaldson-Witten (DW) theory in Wess-Zumino gauge \cite{West:1990}. Such a theory is topological since all its observables are given by metric-independent topological invariants. From the physical point of view, one important motivation to study TQFTs comes from this particular property \textit{i.e.}, the independence with respect to variations on the metric, as it corresponds to a theory that is independent of the background choice. From the mathematical point of view, a TQFT is a very efficient tool to compute topological invariants. For further material on DW theory, see also \cite{Baulieu:1988xs,Brooks:1988jm,Birmingham:1988ap,WerneckdeOliveira:1993pa,Brandhuber:1994uf,Blasi:2000qw,Junqueira:2017zea,Junqueira:2018xgl,Junqueira:2018zxr,Dudal:2019bjh,Junqueira:2021rwc}.

The introduction of a topological phase in gravity would have the power to make a theory of gravity arising from a  background independent topological field theory \cite{Witten:1988ze,Witten:1988xi,vanBaal:1989aw}. Such a phase leads to an intricate problem: How to develop a mechanism to spontaneously break the topological symmetry and release the local degrees of freedom in a TQFT? This is precisely the subject of this paper, where we have demonstrated that the Fujikawa method \cite{Fujikawa:1982ss} could be applied to the twisted $\mathcal{N} = 2$ SYM theory, from which we could build a potential that possesses a nontrivial vacuum. In the end, one obtains a deformed twisted $\mathcal{N} = 2$ SYM theory in the presence of new fields that are introduced in the form of BRST doublets, without explicitly breaking the original symmetries. This allows us to study the spontaneous symmetry breaking (SSB) in a TQFT, and determine how the Higgs mechanism works in the presence of the fermionic scalar supersymmetry described by Witten in \cite{Witten:1988ze}. In short, the supersymmetry is broken together with the Yang-Mills gauge symmetry, and the local degrees of freedom are released, generating mass not only for the vector bosons but also for the fermionic fields of the theory. Although such a method is consistently introduced in DW theory, an actual description for gravity is not constructed in this paper.  We will also not investigate the preservation of Donaldson invariants as the observables of the deformed theory, since the type of topological invariants described by the final Fujikawa-Donaldson-Witten action will not be used to our final result involving the fermionic mass poles in the broken phase. 

Another motivation to study the SSB in DW theory is the connection between topological and conformal field theories (CFT). In three dimensions, it is well-known the correspondence between Chern-Simons theory and a two-dimensional CFT \cite{Witten:1988hf}. In four dimensions, the correlation between TQFTs and the AdS/CFT correspondence was studied, for instance, in \cite{Witten:1998wy, BenettiGenolini:2017zmu}. A wide range of holographic models that seek to describe high-energy physics is based on the AdS/CFT correspondence --- see \cite{PhysRevLett.88.031601, Boschi-Filho:2002wdj, Boschi-Filho:2002xih, PhysRevD.74.015005, Herzog:2006ra, PhysRevD.77.046002, Braga:2022yfe}. The fact that DW theory at the UV regime and Baulieu-Singer gauge theories \cite{Baulieu:1988xs} share the same observables has been repeatedly verified in the literature \cite{Witten:1988ze,Baulieu:1988xs, Weis:1997kj, Delduc:1996yh, Boldo:2003jq}, including the Baulieu-Singer theory in self-dual Landau gauges, which was proved to be conformal \cite{Junqueira:2021rwc}. These examples show the link between TQFTs and CFTs. In Abelian theories, effects involving topological phases of electrons were successfully observed in experiments. Such models describe, for example, phenomena in superfluids \cite{JMKosterlitz_1972, Kosterlitz:1973xp}, quantum Hall conductance \cite{Thouless:1982zz}, and spin chains for antiferromagnets \cite{Haldane:1983ru}. The connection between topology and physics has been successfully verified and opens the possibility of finding topological effects in non-Abelian physics.

The Fujikawa method applied in this topological model contains a prescription that can be used to implement topological phases in field theory, and then perform its spontaneously symmetry breaking (SSB) through the introduction of a well-defined energy scale. The paper is organized as follows. In Section 2, we briefly review the construction of the twisted $\mathcal{N}=2$ SYM action, and its main topological properties. In Section 3, we apply the Fujikawa method to construct a potential with nontrivial vacuum in the bosonic sector, and in Section 4 we analyze the SSB and mass generation for the Higgs mechanism acting together with the fermionic scalar supersymmetry, and compute the fermionic propagators after SSB for a maximal symmetry breaking of the type $SU(3) \rightarrow U(1) \times U(1)$, determining their corresponding mass poles. In Section 5, we discuss the Higgs mechanism and the broken topological phase in a more general way, analyzing the conditions for the release of local degrees of freedom. Section 6 contains our final considerations. 

\section{Topological properties of Donaldson-Witten theory: A brief overview}

 In his seminal paper \cite{Witten:1988ze}, E,~Witten constructed a four-dimensional topological action capable of reproducing the Donaldson invariants \cite{DONALDSON1990257, Donaldson:1983wm, donaldson1987} in the weak coupling limit. Such an action is obtained from a linear transformation of the $\mathcal{N} = 2$ super-Yang-Mills theory, known as twist transformation. 
 
 The multiplet of the $\mathcal{N}=2$ SYM in Wess-Zumino gauge is given by the fields
\begin{equation}
(A_\mu, \psi^i_\alpha, \bar{\psi}^i_{\dot{\alpha}}, \phi, \bar{\phi})\;,
\end{equation}
where $\psi^i_\alpha$ is a Majorana spinor (the super-symmetric partner of the gauge connection $A_\mu$ --- the \textit{gaugino}), and $\phi$, a scalar field. All of these fields belong to the adjoint representation of the gauge group. The indices $(i, \alpha, \dot{\alpha})$ run from one to two, where the index $i$ denotes the internal $SU(2)$ symmetry of the $\mathcal{N}=2$ SYM action, and $(\alpha, \dot{\alpha})$ are Weyl spinor indices: $\alpha$ denotes right-handed spinors, while $\dot{\alpha}$, left-handed ones. The fact that both indices, $i$ and $\alpha$, equally run from one to two suggests the following identification between spinor and supersymmetry indices: 
\begin{equation} \label{identification}
    i \equiv \alpha \;, 
\end{equation}
which defines the twist transformation. The gauge group symmetry of the $\mathcal{N}=2$ SYM action is
\begin{equation}
    SU_L(2) \times SU_R(2) \times SU_I(2) \times U_{R}(1)\;, 
\end{equation}
where $SU_L(2) \times SU_R(2)$ is the rotation group, $SU_I(2)$ is the internal supersymmetry group labeled by $i$, and $U_{{R}}(1)$, the so-called $\mathcal{R}$-symmetry defined by the supercharges ($Q^i_{\alpha}, \,\bar{Q}_{j\dot{\alpha}}$) which are assigned eigenvalues $(+1$, $-1)$, respectively, see \cite{Witten:1988ze}. The identification performed in eq. \eqref{identification} amounts to a modification of the rotation group,
\begin{equation}
SU_L(2) \times SU_R(2) \rightarrow SU_L(2) \times SU_R(2)^\prime\;,
\end{equation}
wherein $SU_R(2)^\prime$ is the diagonal sum of $SU_R(2)$ and $SU_I(2)$. The twisted global symmetry of $\mathcal{N}=2$ SYM takes the form $SU_L(2) \times SU_R(2)^\prime \times U_R(1)$.

In practice, the twist transformation defined by the identification eq. \eqref{identification} only acts on the fermionic fields $(\psi^i_\alpha, \bar{\psi}^i_{\dot{\alpha}})$. The other fields are unaltered. Explicitly, it is given by the linear transformations\footnote{Notation: $\Phi_{(\alpha\beta)} = \Phi_{\alpha\beta} + \Phi_{\beta\alpha}$ and $\Phi_{[\alpha\beta]} =\Phi_{\alpha\beta} - \Phi_{\beta\alpha}$.}
\begin{eqnarray}
\psi^i_\beta \,\,&\rightarrow&\,\, \psi_{\alpha\beta} = \frac{1}{2}\left(\psi_{(\alpha\beta)} + \psi_{[\alpha\beta]}\right) \;, \label{twist1} \\
\bar{\psi}^i_{\dot{\alpha}} &\rightarrow& \bar{\psi}_{\alpha\dot{\alpha}} \,\,\rightarrow\,\, \psi_\mu = (\sigma_\mu)^{\alpha\dot{\alpha}} \bar{\psi}_{\alpha\dot{\alpha}}\,,
\end{eqnarray}
together with
\begin{eqnarray}
\psi_{(\alpha\beta)} \,\, &\rightarrow& \,\, \bar{\chi}_{\mu\nu}  = (\sigma_{\mu\nu})^{\alpha\beta}\psi_{(\alpha\beta)}\;, \\
\psi_{[\alpha\beta]} \,\, &\rightarrow & \bar{\eta}  = \varepsilon^{\alpha\beta} \psi_{[\alpha\beta]}\;,
\end{eqnarray}
where we adopt the conventions for $\varepsilon^{\alpha\beta}$, $(\sigma^\mu)^{\alpha\dot{\alpha}}$ and $(\sigma_{\mu\nu})^{\dot{\alpha}\alpha}$ as the same of \cite{Wess:1992cp}. The field $\bar{\psi}_{\alpha\dot{\alpha}}$ has four independent components as $(\alpha, \dot{\alpha}) = \{1,2\}$, and is mapped into the field $\psi_\mu$. In the other mappings, the same occurs, with $\psi_{(\alpha\beta)}$ mapped into the self-dual field $\bar{\chi}_{\mu\nu}$, while its antisymmetric part, $\psi_{[\alpha\beta]}$, with only one independent component, is mapped into $\bar{\eta}$, a scalar field. We emphasize that $(\psi_\mu, \bar{\chi}_{\mu\nu}, \bar{\eta})$ are anti-commuting fields due to their spinorial origin. 

Since it is a linear transformation, the twist transformation simply corresponds to a change of variables with a trivial Jacobian that could be absorbed in the normalization factor, meaning that both theories (before and after the twist) are perturbatively indistinguishable. In the end, the twist of $\mathcal{N}=2$ SYM action ($S^{N=2}_{SYM}$) \cite{Witten:1988ze, Blasi:2000qw}, in flat Euclidean space, yields the Witten topological Yang-Mills action ($S_{W}$), 
\begin{equation}
S^{N=2}_{SYM}[A_\mu, \psi^i_\alpha, \bar{\psi}^i_{\dot{\alpha}}, \phi, \bar{\phi}] \,\, \rightarrow \,\, S_{W}[A_\mu, \psi_\mu, \bar{\chi}_{\mu\nu}, \bar{\eta}, \bar{\phi}, \phi]\;,
\end{equation} 
where
\begin{eqnarray}\label{SWitten}
S_{W}&=& \frac{1}{g_0^2}Tr\int d^4x \left\{\frac{1}{4} F_{\mu\nu}^+F_{\mu\nu}^+ + \frac{1}{2} \bar{\phi}D_{\mu} D_{\mu} \phi -i\bar{\eta}D_{\mu} \psi_\mu + i D_{\mu} \psi_\nu \cdot \bar{\chi}_{\mu\nu} - \frac{i}{8} \phi \left[ \bar{\chi}_{\mu\nu}, \bar{\chi}_{\mu\nu}\right] +\right.\nonumber\\
&-&\left.\frac{i}{2} \bar{\phi} \left[ \psi_\mu, \psi_\mu \right] - \frac{i}{2} \phi \left[ \bar{\eta}, \bar{\eta}\right] - \frac{1}{8} \left[ \phi, \bar{\phi}\right]^2\right\}\;,
\end{eqnarray}
wherein $F^+_{\mu\nu}$ is the self-dual field\footnote{Following \cite{Witten:1988ze, Blasi:2000qw}, we are considering the positive sign, that corresponds to anti-instantons in the vacuum. A similar construction can be done for instantons, only by changing the sign.}
\begin{equation} 
\label{F^+}
F^+_{\mu\nu} = F_{\mu\nu}+ \widetilde{F}_{\mu\nu}\,, \quad (\widetilde{F}^+_{\mu\nu} = F^+_{\mu\nu})\;,
\end{equation}
with $\widetilde{F}_{\mu\nu} = \frac{1}{2} \epsilon_{\mu\nu\alpha\beta} F_{\alpha \beta}$, and $D_\mu \equiv \partial_\mu - [A_\mu, \,\cdot\,]$ is the covariant derivative in the adjoint representation of the Lie group $G$. The term $\text{Tr}\, F^+_{\mu\nu}F^+_{\mu\nu}$ represents the Yang-Mills action with the Pontryagin boundary. The Witten action \eqref{SWitten} possesses the ordinary Yang-Mills gauge invariance\footnote{It is implicit in this notation the typical text book Yang-Mills transformations of all fields, where the gauge field transforms as $A_\mu^\prime = S^{-1} A_\mu S + S^{-1} \partial_\mu S $ with $S \in G$.},
\begin{equation} \label{gaugesymmetryWitten}
\delta^{\text{YM}}_{\text{gauge}} S_W = 0\;.
\end{equation}
In any case, the theory does not possess gauge anomalies \cite{Maggiore:1994dw}. The symmetry that defines the cohomology of the theory, known as equivariant cohomology, is the fermionic scalar supersymmetry, namely
\begin{eqnarray} \label{Wittensym}
\delta A_\mu &=& -\varepsilon \psi_\mu\,, \quad \delta \phi = 0\,, \quad \delta \bar{\phi} = 2i\varepsilon \bar{\eta}\,, \quad \delta \bar{\eta} = \frac{1}{2}\varepsilon[\phi, \bar{\phi}]\;, \nonumber\\
\delta \psi_\mu &=& - \varepsilon D_\mu \phi\,, \quad \delta \bar{\chi}_{\mu\nu} = \varepsilon F^+_{\mu\nu}\,,
\end{eqnarray}
where $\varepsilon$ is the supersymmetry fermionic parameter that carries no spin, ensuring that the propagating modes of commuting and anti-commuting fields have the
same helicities\footnote{Precisely, the propagating modes of $A_\mu$ have helicities $(1, -1)$, and of $(\phi, \bar{\phi})$, $(0,0)$; while of the fermionic fields $(\bar{\eta}, \psi, \bar{\chi})$, helicities $(1, -1, 0, 0)$.}. The $\delta$-operator must be nilpotent, $\delta^2 = 0$, according to the twisted supersymmetry algebra. This symmetry relates bosonic and fermionic degrees of freedom, which are identical --- an inheritance of supersymmetry\footnote{The action $S_W$ is also invariant under global scaling with dimensions $(1, 0, 2,2,1,2)$ for $(A, \phi, \bar{\phi}, \bar{\eta}, \psi, \bar{\chi})$, respectively; and preserves an additive $U$ symmetry for the assignments $(0, 2, -2, -1,1, -1)$. In the BRST formalism, the latter is naturally recognized as ghost numbers.}. As we are working in the Wess-Zumino gauge, one needs to use the equations of motion to recover the nilpotency of $\delta$ \cite{Blasi:1989ka, Witten:1988ze}. 
Such a symmetry is, then, associated to an on-shell nilpotent ``BRST charge", $\mathcal{Q}$, according to the definition of the $\delta$ variation of any functional $\mathcal{O}$ as a transformation on the space of all functionals of field variables, namely, 
\begin{equation} \label{deltaQ}
\delta \mathcal{O} = -i\varepsilon \cdot \{\mathcal{Q}, \mathcal{O}\}\,, \quad \text{such that} \quad \mathcal{Q}^2\vert_{\textit{on-shell}} = 0\;,
\end{equation} 
after using the $\bar{\chi}$ equation of motion.

This supersymmetry is also valid for an arbitrary orientable Riemannian four-manifold, meaning  that one can perform 
\begin{equation}\label{anyR}
    \int d^4x \rightarrow \int d^4x \sqrt{-g}
\end{equation}
in Witten action \eqref{SWitten}. In the computation of the $\delta$-symmetry, the commutator $\left[D_\mu, D_\nu\right]$ only appears acting on the spin zero scalar field $\phi$, so that the Riemannian tensor does not appear in the verification of the supersymmetry, proving its validation in curved spaces.  The action \eqref{SWitten} has the property of being invariant under infinitesimal changes in the metric. This property characterizes the Witten model as a topological quantum field theory, and can be verified by the fact that the energy-momentum tensor, $T_{\mu\nu}$, is a BRST-exact term, see \cite{Witten:1988ze},
\begin{equation} \label{deltaTexact}
T_{\mu\nu}^{(W)} =  \{ \mathcal{Q}, V_{\mu\nu}^{(W)}\}\,, \quad \delta^2\vert_{\textit{on-shell}}=0\;,
\end{equation}
with
\begin{eqnarray}\label{Vmunu}
V_{\mu\nu}^{(W)} &=&  \frac{1}{2} \text{Tr}\{ F_{\mu\sigma} \bar{\chi}_\nu^{\,\,\,\sigma} + F_{\nu\sigma}\bar{\chi}_\mu^{\,\,\,\sigma} - \frac{1}{2} g_{\mu\nu}F_{\sigma\rho}\bar{\chi}^{\sigma\rho}\} + \frac{1}{4}g_{\mu\nu} \text{Tr} \bar{\eta} [\phi, \bar{\phi}] \nonumber\\
 &+&\frac{1}{2} \text{Tr} \{ \psi_\mu D^\nu \bar{\phi} + \psi_\nu D^\mu \bar{\phi} - g_{\mu\nu}\psi_{\sigma} D^{\sigma}\bar{\phi}  \} \;.
\end{eqnarray}

Equation \eqref{deltaTexact} together with $\delta S_W = 0$ shows that Witten theory is a topological quantum field theory, since \begin{eqnarray}\label{gmunuinv}
\frac{\delta}{\delta g^{\mu\nu}} Z &=& \int \mathcal{D}\Phi (- \frac{\delta}{\delta g^{\mu\nu}} \mathcal{S}_W)\text{exp}(-\mathcal{S}_W) \nonumber\\
&=& - \frac{1}{g_0^2} \langle \{ \mathcal{Q}, \int_M d^4x \sqrt{g}  V_{\mu\nu}^{(W)}\}\rangle = 0\;,
\end{eqnarray}
as all expectation values of a BRST-exact term vanish, where $Z$ is the partition functional. As it is well known, the Feynman path integral of Witten TQFT naturally reproduces the Donaldson invariants for four-manifolds, that represent the global observables of the theory. In general, a topological quantum field theory is defined by these two conditions: (i) $\delta S = 0$; (ii) $T_{\mu\nu}$ is $\delta$-exact, 
which are sufficient to state that the observables are independent of variations on the metric. The type of topological invariants described as observables of the theory depends on the model. 

Our aim is to construct a potential that possesses a nontrivial vacuum, by introducing an extra term to Witten action, without destroying its topological properties. Such a potential would promote the spontaneous breaking of the gauge symmetry, thus releasing the local degrees of freedom via a Higgs-like mechanism. As we will see, this can be done by applying the Fujikawa symmetry breaking method \cite{Fujikawa:1982ss}.

\section{Fujikawa method: Construction of a potential with nontrivial vacuum in the trivial part of equivariant cohomology}

The construction of a potential via Fujikawa's \cite{Fujikawa:1982ss} method consists of introducing a BRST-exact Higgs-like potential \cite{Higgs:1964pj} into the original action. In our case, it should be a $\delta$-exact term, since it is the $\delta$-operator that defines the cohomology of Witten's theory. Such a term will not break the topological properties of Witten action, as the energy-momentum tensor will remain a BRST-exact term, with the $\delta$-symmetry preserved due to the nilpotency of the $\delta$-operator. For that, the new term should not modify the $\bar{\chi}_{\mu\nu}$ equation of motion. Such a term will not affect the topological nature of the observables, which will continue to be invariant under metric variations, since the addition of a BRST-exact term to Witten's action would only generate in the Feynman path integral a contribution that corresponds to a expectation value of a $\delta$-exact term. 

In order to construct a potential with nontrivial vacuum via Fujikawa's method, one must add the following term to the action in the trivial part of $\delta$-cohomology\footnote{We are using the notation $[\Psi, \Phi]^a = f^{abc} \Psi^b \Phi^c$.}:
\begin{eqnarray}\label{Sssb}
S_{F}&=& \frac{1}{g_0^2} \int d^4x\left[\xi^a \bar{\eta}^a \left(\bar{\phi}^b \phi^b - v^2\right) +\Xi^a \left[ \phi, \bar{\phi} \right]^a \bar{\phi}^b \phi^b - 2i \Xi^a \bar{\eta}^a\bar{\eta}^b \phi^b - v^2\Xi^a \left[ \phi, \bar{\phi} \right]^a\right.\nonumber\\
 &+& \left. \bar{\theta}^a \Upsilon^b \left( \phi^a \Xi^b + \frac{1}{2} f^{abc}\phi^c \right) + \bar{\Theta}^a \phi^a \Upsilon^b \xi^b + \zeta^a \Upsilon^b \left( \bar{\phi}^a \Xi^b - \frac{1}{2} f^{abc}\bar{\phi}^c \right) \right.\nonumber\\ &+& \left. 2iZ^a \Upsilon^b \left(  \bar{\eta}^{a}\Xi^b - \frac{1}{2} f^{abc}\bar{\eta}^c \right) + Z^a\bar{\phi}^a \Upsilon^b \xi^b  \right]\;,
\end{eqnarray}
where $\left(\Xi^a,\, \xi^a, \,\bar{\Theta}^a,\, \bar{\theta}^a,\, Z^a, \zeta^a, \Upsilon^a\right)(x)$ are Lorentz scalar fields in the adjoint representation of the gauge group with ghost numbers $\left(0,1,-2,-3,2,1,0\right)$, respectively, being $\Upsilon^a(x)$ a bosonic field, with 
\begin{equation}\label{deltaUpsilon}
\delta \Upsilon^a = 0\;,
\end{equation}
and with the $\delta$-transformations of the other fields defined as
\begin{eqnarray}
 \delta \Xi^a &=& \varepsilon \xi^a\;, \label{delta1}\\
 \delta \xi^a &=& 0\;,\label{delta2}\\
 \delta \bar{\Theta}^a &=& \varepsilon \bar{\theta}^a\;,\\
 \delta \bar{\theta}^a &=& 0\;, \\
 \delta Z^a &=& \varepsilon \zeta^a\;,\\
 \delta \zeta^a &=& 0\;.\label{delta6}
\end{eqnarray}
Using the transformations \eqref{delta1}-\eqref{delta6}, the Fujikawa's additional action can be written as a $\delta$-exact term, $S_{F} = \delta W_{F}$, with\footnote{This is not the same $\delta$-exact term introduced by K. Fujikawa in \cite{Fujikawa:1982ss}, which contains massive fermions. As the method is the same, based on the introduction of an energy scale in the trivial part of $\delta$-cohomology, we will call $S_F$ the Fujikawa-type action, or simply, Fujikawa action. } 
\begin{eqnarray}
    W_{F} &=& \frac{1}{g_0^2}\int d^4x\bar{\varepsilon}  \left[ \Xi^a \eta^a \left(\bar{\phi}^b \phi^b - v^2\right)  + \bar{\Theta}^a \Upsilon^b\left( \phi^a \Xi^b + \frac{1}{2} f^{abc}\phi^c \right)+Z^a \Upsilon^b\left( \bar{\phi}^a \Xi^b - \frac{1}{2} f^{abc}\bar{\phi}^c \right)\right]\;, 
\end{eqnarray}
where $\bar{\varepsilon} \varepsilon = 1$, by definition. 

The new $\delta$-transformations defines the total fermionic scalar supersymmetry in the form
\begin{equation}\label{deltaT}
 \delta_T = \delta + \delta_F\;,    
\end{equation}
with 
\begin{equation}
    \delta_F = \int d^4x\left( \varepsilon \xi^a \frac{\delta}{\delta \Xi^a } + \varepsilon \bar{\theta}^a \frac{\delta}{\delta \bar{\Theta}^a}+\varepsilon \zeta^a \frac{\delta}{\delta Z^a}\right) \;.
\end{equation}
The $\delta_T$-operator, which defines the global observables in the presence of the new fields, is automatically  nilpotent due to BRST nature of the new fields, which are given by 3 BRST doublets and one BRST-invariant field ($\Upsilon$). As $\delta_F$ only acts on the new fields, and $\delta$ on the original ones, one has
\begin{eqnarray}
    \delta_T \Phi_{\text{new fields}} &=& \delta_F \Phi_{\text{new fields}}\;, \\
    \delta_T \Phi_{\text{Witten fields}} &=& \delta \Phi_{\text{Witten fields}}\;,
\end{eqnarray}
such that
  \begin{equation}
    \delta^2_T\Phi_{\text{all fields}} = 0\;,  
\end{equation}  
since $\delta^2_F \Phi_{\text{new fields}} = 0$, without requiring any equation of motion. The $\delta$-operator is still on-shell nilpotent, using the $\bar{\chi}_{\mu\nu}$ equation of motion, that was preserved in the presence of the new fields, once $S_F$ does not depend on $\bar{\chi}_{\mu\nu}$.

For the sake of simplicity, we consider a Euclidean flat spacetime for the calculations in this section, remembering that we can always use \eqref{anyR} since the Riemannian tensor does not appear in the verification of the scalar $\delta$-supersymmetry in the same way as before. The new fields $(\Xi, \xi)$ and $(\bar{\Theta}, \bar{\theta})$ introduced here form two BRST doublets, that do affect the local description of the observables, see \cite{Dixon:1991wi,Piguet:1995er,Vandersickel:2011zc}, and that automatically preserves the nilpotency of the $\delta$-operator. Naturally, $v$ is a mass scale indispensable for a SSB mechanism. The bosonic Witten potential alone,
\begin{equation}\label{Trphi2}
\text{Tr}\left[ \phi, \bar{\phi}\right]^2\;,
\end{equation}
does not fix an energy scale in the vacuum. In fact, it does not break the fermionic scalar supersymmetry. For example, in a $SU(2)$ gauge theory, the vacuum solution of \eqref{Trphi2}, $\phi=\frac{1}{2} \alpha \sigma^3$ with $\alpha$ an arbitrary complex parameter, is invariant under $\delta$-transformations.    

We would like to emphasize that to construct a potential with $(\phi, \bar{\phi})$ fields without other bosonic mixtures we needed to explore the $\delta$-transformation of the $\bar{\eta}$ field, see \eqref{Wittensym}. At this point, we focus on the bosonic sector $(\bar{\phi}, \phi, \Xi)$ of the potential, namely,
\begin{equation}\label{Ussb}
U_{SSB}(\bar{\phi}, \phi, \Xi) = \Xi^a \left[ \phi, \bar{\phi} \right]^a \bar{\phi}^b \phi^b - v^2 \Xi^a \left[ \phi, \bar{\phi} \right]^a + \frac{1}{4} \left(\left[\phi, \bar{\phi} \right]^a\right)^2  \;,  
\end{equation}
wherein the last term comes from the original Witten action, and the first two, from the $\delta$-exact Fujikawa term $S_{F}$, see Eq. \eqref{Sssb}. The minimum of the potential \eqref{Ussb} obeys 
\begin{eqnarray}
 \frac{\partial U_{SSB}(\bar{\phi}, \phi, \Xi)}{\partial \bar{\phi}^j} &=& 0 \;,\label{minU1} \\
 \frac{\partial U_{SSB}(\bar{\phi}, \phi, \Xi)}{\partial \phi^j} &=& 0\;, \\ 
 \frac{\partial U_{SSB}(\bar{\phi}, \phi, \Xi)}{\partial \Xi^j} &=& 0 \;,\label{minU3} 
\end{eqnarray}
which yields, respectively,
\begin{eqnarray}
\left[\Xi,\bar{\phi}\right]^j\left(\bar{\phi}^b \phi^b - v^2\right) + \left[ \phi, \bar{\phi} \right]^a  \left( \phi^j\Xi^a + \frac{1}{2} f^{jab} \phi^b\right) \vert_{vacuum}&=& 0\;,\label{v1} \\
-\left[\Xi,\bar{\phi}\right]^j\left(\bar{\phi}^b \phi^b - v^2\right) + \left[ \phi, \bar{\phi} \right]^a \left( \bar{\phi}^j \Xi^a  - \frac{1}{2} f^{jab}  \bar{\phi}^b\right) \vert_{vacuum}&=& 0\;,\label{v2}\\
  \left[ \phi, \bar{\phi} \right]^j \left(\bar{\phi}^b \phi^b - v^2  \right) \vert_{vacuum}&=& 0\;.\label{v3}
 \end{eqnarray}
Assuming the invariance of the vacuum under Lorentz transformations, and according to the ghost numbers of the scalar fields, the condition
\begin{equation}\label{vacuum1}
    \bar{\phi}^b_0  \phi^b_0 \equiv \bar{\phi}^b \phi^b \vert_{vacuum}  =  v^2\;, 
\end{equation}
with
\begin{eqnarray}\label{vacuum2}
\phi^a_0(x) &=& \kappa v^a(x) +\rho\bar{\eta}^a(x)\nonumber\\ \bar{\phi}^a_0(x) &=& \bar{\kappa} v^a(x) + \bar{\rho} \bar{\eta}^a(x) \;, 
\end{eqnarray}
where $v^a(x)$ is a scalar field with ghost number zero satisfying $v^av^a = v^2$ in the vacuum, and $(\kappa, \bar{\kappa},\rho, \bar{\rho})$ are parameters with ghost numbers $\left(2,-2,3,-1\right)$, obeying 
\begin{eqnarray}\label{cond}
\kappa \bar{\kappa} = 1\;,\quad
\kappa \bar{\rho} + \rho \bar{\kappa} = 0\;,
\end{eqnarray}
is a possible solution of equations \eqref{v1}- \eqref{v3} for $\Xi^a(x)\vert_{vacuum}$ not necessarily zero, which shows that Witten-Fujikawa (WF) action can possess nontrivial vacuum with a well-defined $v$. We can also consider this problem for operators $(\hat{\kappa},\hat{\bar{\kappa}})$ and $(\hat{\rho},\hat{\bar{\rho}})$ acting on the fields $v^a(x)$ and $\bar{\eta}^a(x)$, being $(\kappa,\bar{\kappa})$ and $(\rho, \bar{\rho})$ their eigenvalues, respectively, with the operators obeying the \footnote{We do not consider $\delta_T$ in this case, since $\delta_T \phi_0 = \delta \phi_0$ and $\delta_T \bar{\phi}_0 = \delta \bar{\phi}_0$. }
\begin{eqnarray}\label{algebra}
\hat{\kappa} \hat{\bar{\kappa}} = 1\;,\quad
\hat{\kappa} \hat{\bar{\rho}} + \hat{\rho} \hat{\bar{\kappa}} &=& 0\;, \nonumber\\
\left[ \hat{\kappa},\delta \right] =\left[\hat{\bar{\kappa}},\delta \right] &=& 0\;,\nonumber\\
\left[ \hat{\rho}, \delta \right] = \left[\hat{\bar{\rho}}, \delta \right] &=& 0\;. 
\end{eqnarray}
From the vacuum solution \eqref{vacuum2},
\begin{equation}\label{[phi,phi]0}
    \left[\phi_0, \bar{\phi}_0\right]^a = (\kappa \bar{\rho} -\rho \bar{\kappa})f^{abc}v^b\bar{\eta}^c + \rho\bar{\rho}f^{abc}\bar{\eta}^b\bar{\eta}^c\;, 
\end{equation}
which must be null in the vacuum, due to the condition for the solution $(\phi_0, \bar{\phi}_0)$ to represent a minimum of the potential, according to the SSB mechanism. The second and third terms of the equations \eqref{v1} and \eqref{v2} vanish using the  $\bar{\theta}$- and $\zeta$-equations of motion, assuming the solution where $\Upsilon$ is not necessarily zero ($\Upsilon \neq 0$). This way,  Eq. \eqref{[phi,phi]0} will be consistent in the vacuum if 
\begin{equation}\label{[minimum}
   (\kappa \bar{\rho} -\rho \bar{\kappa})f^{abc}v^b\bar{\eta}^c + \rho\bar{\rho}f^{abc}\bar{\eta}^b\bar{\eta}^c = 0\;, 
\end{equation}
such that, taking the $\delta$-transformation of the equation, and using $\delta \bar{\eta}^a\vert_{\text{vacuum}} = \frac{1}{2}\varepsilon [\phi_0, \bar{\phi}_0] = 0$, 
\begin{equation}\label{minimum}
   (\kappa \bar{\rho} -\rho \bar{\kappa})f^{abc}\delta v^b\bar{\eta}^c = 0\;,
\end{equation}
which possess a nontrivial solution\footnote{To obtain a non-trivial solution, we must consider the condition  $2 \bar{\kappa} v^a \neq  \bar{\rho}\bar{\eta}^a$.}, with $\kappa \bar{\rho} \neq\rho \bar{\kappa}$, if we define the $\delta$-transformation of $v^a$ as
\begin{eqnarray}\label{deltav}
    \delta v^a = \varepsilon \alpha f^{abc} \Upsilon^b \xi^c\;,
\end{eqnarray}
being $\alpha$ a parameter with ghost number 0, and assuming that $\Upsilon$ and $\xi$ do not belong to the Cartan's subalgebra, with $\xi$ containing non-zero components in the vacuum, obeying\footnote{This condition is in harmony with $\xi^a\vert_{vacuum} \neq 0$. The derivatives with respect to the fields of the potential $V_0(\xi, \eta, \phi, \bar{\phi})$, see Eq. \eqref{Sssb}, are naturally satisfied in the vacuum, \textit{i. e.} $\partial_\xi V_0 =  \partial_\eta V_0 =  \partial_\phi V_0 =  \partial_{\bar{\phi}} V_0 = 0$, using Eq. \eqref{vacuum1} and the condition \eqref{Yn}, with $\xi^a$ $\forall a$ being able to assume non-zero values in the vacuum, without affecting the local minimum of the potential $U_{SSB}$ in the subspace $(\phi, \bar{\phi})$, since the vacuum solution \eqref{vacuum1} does not depend on $\xi$.}
\begin{eqnarray}\label{Yn}
    \Upsilon^a \bar{\eta}^a\vert_{\text{vacuum}} = 0 \quad \text{and} \quad \xi^a \bar{\eta}^a\vert_{\text{vacuum}} = 0\;.
\end{eqnarray}
One observes that the $\delta$-transformation \eqref{deltav} is off-shell nilpotent, according to the $\delta$-transformations of $\Upsilon$ and $\xi$, see equations \eqref{deltaUpsilon} and \eqref{delta2}. The mass dimension of $\alpha$ is zero, and it does not contribute to the energy scale generated via SSB, since it does not appear in the vacuum solution \eqref{vacuum1}.

Using Eq. \eqref{deltav},  the $\delta$-transformation of \eqref{vacuum2} in the vacuum yields
\begin{eqnarray}
\delta \phi^a_0 &=& \varepsilon\alpha \kappa f^{abc} \Upsilon^b \xi^c \vert_{\text{vacuum}}\;, \label{deltaphi01}\\
\delta \bar{\phi}^a_0 &=& \varepsilon \alpha \bar{\kappa} f^{abc} \Upsilon^b \xi^c \vert_{\text{vacuum}}\;,\label{deltaphi02}
\end{eqnarray} 
then, in this case, this vacuum solution will be able to also spontaneously break the fermionic scalar supersymmetry via the Higgs mechanism, since $\phi_0$ and $\bar{\phi}_0$ are shown to be not invariant under $\delta$-transformations\footnote{To break the supersymmetry, we need 3 different directions of $\phi$ in the vacuum: one for $v^a$ and two for $\bar{\eta}^a$. If we choose only one direction for $\bar{\eta}^a$ in the vacuum, namely $a = J$, Eq. \eqref{Yn} would imply $\Upsilon^J \bar{\eta}^J\vert_{\text{vacuum}} = 0$. As $\Upsilon^a$ is non-null in the non-trivial vacuum solution, we would get $\bar{\eta}^J = 0$, necessarily, which would trivialize Eq. \eqref{minimum} that fixes the $\delta$-transformation of $v^a$.}.

\section{Broken topological phase, SSB and mass generation for bosons and fermions}\label{MASS}

The solution for the minimum of the potential $U_{SSB}$ given by equations \eqref{vacuum1} and \eqref{vacuum2} indicates that the potential \eqref{Ussb} is able to promote the spontaneous breaking of gauge symmetry, thus generating mass for bosonic vector fields. Together with the YM gauge symmetry, the fermionic scalar supersymmetry \eqref{Wittensym} that characterizes the topological theory will also be broken in this process, implying that the potential \eqref{Ussb} has the power of releasing the local degrees of freedom, that are ``confined" in the topological phase (before the SSB), which only possess global observables. The break of the $\delta$-symmetry will also be responsible, via the Higgs mechanism, for generating mass for the fermionic fields $\bar{\eta}^a$, $\bar{\chi}^a_{\mu\nu}$ and $\psi^a_\mu$.

The spontaneous symmetry breaking can occur when a particular direction of the degenerate vacuum is chosen, that is not invariant under symmetry transformations of the original action. In particular, for a theory with color symmetry defined by $G = SU(N)$, choosing the vacuum directions 
\begin{equation}\label{na}
v^a\vert_{\text{vacuum}} = \begin{pmatrix}
0 \\
0 \\
\vdots\\
0\\
v
\end{pmatrix}\;, \quad \bar{\eta}^a\vert_{\text{vacuum}}  = \begin{pmatrix}
0 \\
\vdots \\
H_1\\
H_2\\
0
\end{pmatrix}\;,
\end{equation}
with $a = \{1, \cdots, D\}$, being $D = N^2-1$ the $SU(N)$ dimension. In this notation, $H_1 = \bar{\eta}^{D-2}$ and $H_2 = \bar{\eta}^{D-1}$ in the vacuum, which allows to promote the symmetry breaking of the scalar supersymmetry, since we need 3 different directions in the vacuum involving the fields $(v^a, \bar{\eta}^a)$, in order to obtain a vacuum that breaks the topological phase, see equation \eqref{deltaphi01} and \eqref{deltaphi02}. As mentioned in the previous section, to be a minimum of the potential, $\phi_0$ and $\bar{\phi_0}$ must commute. The off-diagonal terms give positive contributions to the potential energy. In other words, $\phi_0$ must belong to the subalgebra of Cartan. For the directions \eqref{na}, one must have 
\begin{eqnarray}
    \phi_0  = \phi_0^{i} \frac{T^i_3}{2}\;, \nonumber\\
    \bar{\phi}_0 = \bar{\phi}_0^{i}  \frac{T^i_3}{2}\;,
\end{eqnarray}
with $i = \{D-2, D-1, D\}$, and $[T_3^i, T_3^j] = 0$. For this reason, the scalar supersymmetry can only be broken for gauge groups that have 3 or more generators in the Cartan subalgebra. Replacing \eqref{na} into \eqref{vacuum2}, one has
\begin{eqnarray}
(\phi_0^D, \phi_0^{D-1}, \phi_0^{D-2}) &=& (\kappa v, \rho H_1, \rho H_2)\,\\
(\bar{\phi}_0^D, \bar{\phi}_0^{D-1}, \bar{\phi}_0^{D-2}) &=& (\bar{\kappa} v, \bar{\rho} H_1, \bar{\rho} H_2)\,.
\end{eqnarray}

In the broken phase, the bosonic scalar fields for the vaccum directions \eqref{na} take the form:
\begin{eqnarray}
    \phi^a\vert_{\text{after SSB}} = \sum_{i=1}^{D-3} \phi^i \delta^{ai}+(\omega_1 + v_{\rho 1})\delta^{a(D-2)} + (\omega_2 + v_{\rho 2})\delta^{a(D-1)}+ (\omega + v_\kappa)\delta^{aD}\;,\label{phiSSB}\\
     \bar{\phi}^a\vert_{\text{after SSB}} = \sum_{i=1}^{N-3} \bar{\phi}^i \delta^{ai}+(\bar{\omega}_1 + \bar{v}_{\rho 1})\delta^{a(N-2)} + (\bar{\omega}_2 + \bar{v}_{\rho 2})\delta^{a(N-1)}+ (\bar{\omega} + \bar{v}_\kappa)\delta^{aN}\;,\label{barphiSSB}
\end{eqnarray}
wherein $\kappa v \equiv v_\kappa$, $\bar{\kappa} v \equiv \bar{v}_\kappa$, $v_{\rho\, 1} \equiv \rho H_1$, $v_{\rho\, 2}$ and $\equiv \rho H_2$. One identifies the scalar fields $(\omega, \bar{\omega}$), $(\omega_1, \bar{\omega}_1$) and $(\omega_2, \bar{\omega}_2)$  as the fluctuations around the vacuum  for the $D$, $D-2$ and $D-1$ components of  $\phi^a$ and $\bar{\phi}^a$ after SSB, respectively. 

Using the $D$-components of $\phi^a$ and $\bar{\phi}^a$, see equations \eqref{phiSSB} and \eqref{barphiSSB}, one gets
\begin{eqnarray}\label{Dphi}
D_\mu^{ab} \phi^b
= \partial_\mu \omega \delta^{aD} - A^{aD}_\mu (\omega + v_\kappa)\;, 
\end{eqnarray}
where we have defined 
\begin{equation}\label{defA}
    A_\mu^{aD} = f^{aDb}A_\mu^b\;. 
\end{equation}
From equation \eqref{Dphi} and the definition above, one gets
\begin{eqnarray}
D_\mu^{ac} D_\mu^{cd} \phi^d &=& \partial^2 \omega \delta^{aD} - \partial_\mu A_\mu^{aD}(\omega+v_\kappa) - 2 A_\mu^{aD} \partial_\mu \omega -f^{acd}A_\mu^{cD} A_\mu^d (\omega + v_\kappa)\;,
\end{eqnarray}
which is used to obtain the following bosonic term of Witten's action after SSB:
\begin{eqnarray} \label{phiDDphi}
\bar{\phi}^a D_\mu^{ac} D_\mu^{cd} \phi^d &=& \bar{\omega} \partial^2 \omega - v^2 A_\mu^{aD}A_\mu^{aD} - v_\kappa \bar{\omega} A_\mu^{aD}A_\mu^{aD} - \bar{v}_\kappa \omega A_\mu^{aD}A_\mu^{aD}- \bar{\omega} \omega A_\mu^{aD}A_\mu^{aD}\;,
\end{eqnarray}
where $\bar{v}_\kappa v_\kappa = v^2$ according to \eqref{cond}.

The term above displays one clear difference when compared with the SSB in the ordinary Yang-Mills case. Here, the fields $(\bar{\omega},\omega)$ that represent the oscillations around the nontrivial vacuum do not possess a quadratic mass term. This occurred because the quadratic term in the potential involving the fields $\bar{\phi}$ and $\phi$ is proportional to the commutator $\text{Tr}\,[\bar{\phi}, \phi]$ instead of $\text{Tr}\, \bar{\phi} \phi$, see Eq. \eqref{Ussb}. This commutator has appeared in the Fujikawa term of potential, being naturally introduced via the transformation of the field $\bar{\eta}$ under the fermionic scalar supersymmetry, see \eqref{Wittensym}, thus being protected by the preservation of equivariant cohomology in the topological case. On the other hand, the second term of Eq. \eqref{phiDDphi} is similar to the ordinary Yang-Mills term that comes from the Higgs mechanism. One must note that $A_\mu^{ab} = 0$ if $a = b$ due to the antisymmetric property of $f^{abc}$ -- see definition \eqref{defA}. In ordinary YM, these massive bosons ``absorb" the Goldstone bosons that disappear from the spectrum. 

The situation is different in the topological case, since the vacuum must possess, at least, 3 different directions. In an effective way, however, the Higgs mechanism will still work. To compute the SSB effect due to the components $\phi^{D-2}$ and $\phi^{D-1}$ with nontrivial vacuum, we should use again equations \eqref{phiSSB} and \eqref{barphiSSB}. By replacing these components into $\text{Tr} \bar{\phi} D_\mu D_\mu \phi$, we would obtain the same equation \eqref{phiDDphi} for the corresponding fluctuations $(\omega_i, \bar{\omega}_i)$, but without the massive term for $A_\mu^{a (D-1)}$ or $A_\mu^{a (D-2)}$. The reason is due to the fermionic nature of $H_1$ and $H_2$, such that $\bar{v}_{\rho\, 1} v_{\rho\, 1} =\bar{v}_{\rho\, 2} v_{\rho\, 2}= 0$. The mass generation for the vector gauge fields is entirely determined by the $D$-components of $\bar{\phi}^a$ and $\phi^a$ for the vacuum with direction $\delta^{a D}$. The SSB also generates extra quadratic and interacting terms to the action from the components oscillating around the nontrivial vacuum.

The Witten's action sector corresponding to the components of $\phi^a$ and $\bar{\phi}^a$ with nontrivial vacuum, using the last 3 terms of equations \eqref{phiSSB} and \eqref{barphiSSB},  becomes
\begin{eqnarray}\label{SWafterSSB}
S_W\vert_{\text{after SSB}} &=& \frac{1}{2g^2} \int d^4x \left[\frac{1}{4} F^{a+}_{\mu\nu} F^{a+}_{\mu\nu} + \frac{1}{2}\left( \bar{\omega}_I \partial^2 \omega_I - v^2 A_\mu^{aD}A_\mu^{aD}\right) -i\bar{\eta}^b D_\mu^{bc} \psi^c_\mu + i D_\mu^{bc} \psi^c_\nu \cdot \bar{\chi}^b_{\mu\nu}\right. \nonumber\\
&+&\left. \frac{v_\kappa}{8} f^{Dbc} \bar{\chi}^b_{\mu\nu}\bar{\chi}^c_{\mu\nu}  +\frac{\bar{v}_\kappa}{2}f^{Dbc} \psi^b_\mu \psi^c_\mu + \frac{v_\kappa}{2} f^{Dbc} \bar{\eta}^b\bar{\eta}^c  \right. \nonumber\\
&+&\left. \frac{v_{\rho \, 1}}{8} f^{(D-2)bc} \bar{\chi}^b_{\mu\nu}\bar{\chi}^c_{\mu\nu}  +\frac{\bar{v}_{\rho \,1}}{2}f^{(D-2)bc} \psi^b_\mu \psi^c_\mu + \frac{v_{\rho \, 1}}{2} f^{(D-2)bc} \bar{\eta}^b\bar{\eta}^c \right. \nonumber\\
&+&\left. \frac{v_{\rho \, 2}}{8} f^{(D-1)bc} \bar{\chi}^b_{\mu\nu}\bar{\chi}^c_{\mu\nu}  +\frac{\bar{v}_{\rho \,2}}{2}f^{(D-1)bc} \psi^b_\mu \psi^c_\mu + \frac{v_{\rho \, 2}}{2} f^{(D-1)bc} \bar{\eta}^b\bar{\eta}^c + \mathcal{L}_1 +\mathcal{L}_2 \right]  \;, 
\end{eqnarray}
with $\omega_I \equiv (\omega, \omega_1, \omega_2)$, wherein $\mathcal{L}_1[\omega_I, \bar{\omega}_I, \psi, \bar{\chi}, \bar{\eta}]$  represents the following interactions
\begin{eqnarray}
   \mathcal{L}_1 &=& \frac{1}{2}\left( \bar{v}_\kappa \omega A_\mu^{aD}A_\mu^{aD}- \bar{\omega} \omega A_\mu^{aD}A_\mu^{aD}+\bar{v}_{\rho1} \omega_1 A_\mu^{a(D-1)}A_\mu^{a(D-1)}- \bar{\omega}_1 \omega_1 A_\mu^{a(D-1)}A_\mu^{a(D-1)} \right. \nonumber\\
   &+& \left.\bar{v}_{\rho2} \omega_2 A_\mu^{a(D-2)}A_\mu^{a(D-2)}- \bar{\omega}_2 \omega_2 A_\mu^{a(D-2)}A_\mu^{a(D-2)}\right)\nonumber\\
   &+&\frac{1}{8}f^{Dbc} \omega \bar{\chi}^b_{\mu\nu} \bar{\chi}^c_{\mu\nu} +  \frac{1}{2}f^{Dbc} \bar{\omega} \psi^b_{\mu} \psi^c_{\mu}+  \frac{1}{2}f^{Dbc} \omega \bar{\eta}^b \bar{\eta}^c\nonumber\\
   &+& \frac{1}{8}f^{(D-2)bc} \omega_1 \bar{\chi}^b_{\mu\nu} \bar{\chi}^c_{\mu\nu} +  \frac{1}{2}f^{(D-2)bc} \bar{\omega}_1 \psi^b_{\mu} \psi^c_{\mu}+  \frac{1}{2}f^{(D-2)bc} \omega_1 \bar{\eta}^b \bar{\eta}^c\nonumber\\
   &+& \frac{1}{8}f^{(D-1)bc} \omega_2 \bar{\chi}^b_{\mu\nu} \bar{\chi}^c_{\mu\nu} +  \frac{1}{2}f^{(D-1)bc} \bar{\omega}_2 \psi^b_{\mu} \psi^c_{\mu}+  \frac{1}{2}f^{(D-1)bc} \omega_2 \bar{\eta}^b \bar{\eta}^c\;,
\end{eqnarray}
while 
\begin{eqnarray}
\mathcal{L}_2[\bar{\omega}, \omega] &=& \frac{1}{4} \left[f^{abc}(\omega + v_\kappa)\delta^{bD} (\bar{\omega}_1 + \bar{v}_{\rho 1})\delta^{c (D-1)}+f^{abc}(\omega + v_\kappa)\delta^{bD} (\bar{\omega}_2 + \bar{v}_{\rho 2})\delta^{c (D-2)}\right.\nonumber\\
&+&\left. f^{abc}(\bar{\omega}_1 + v_{\rho 2})\delta^{b(D-1)} (\bar{\omega}_2 + \bar{v}_{\rho 2})\delta^{c (D-2)}\right]^2 \;,
\end{eqnarray}
with $D=8$ for the $G=SU(3)$ case.

One observes in Eq. \eqref{SWafterSSB} the appearance of quadratic terms for the fermionic fields $\bar{\chi}_{\mu\nu}$, $\psi_\mu$ and $\bar{\eta}$, suggesting that these fields have gained mass. There is a mixture between them, due to the terms $\bar{\eta} D_\mu \psi$ and $D_\mu \psi_\nu \cdot \bar{\chi}^c_{\mu\nu}$. Such a mixture opens the possibility of mass generation for the fields $\bar{\chi}_{\mu\nu}$, $\psi_\mu$ and $\bar{\eta}$, with their masses written in terms of the energy scale introduced by the Fujikawa potential. This mass generation can be understood through the Higgs mechanism, for a spontaneous breaking of the gauge symmetry together with the fermionic scalar supersymmetry that correlates the bosonic and fermionic degrees of freedom.

\subsection{Higgs mechanism for $SU(3) \rightarrow U(1) \times U(1)$ in the topological case: Mass generation for bosons and fermions}

In order to exemplify how the Higgs mechanism actually works in the topological Yang-Mills case, we will analyze the symmetry breaking $SU(3) \rightarrow U(1) \times U(1)$ assuming that the Witten-Fujikawa action has a starting $G = SU(3)$ color symmetry. This is a possible choice to break the supersymmetry, since this group has 3 generators in the Cartan subalgebra. Assuming the vacuum directions \eqref{na} with distinct values for $\phi_0^i$, representing a maximal SSB, the vacuum is invariant under transformations defined by a matrix of the type $\hat{R}_{5\times 5}$ such that
\begin{eqnarray}
    [\hat{R}, \phi_0] = 0\;.
\end{eqnarray}
which implies that the resulting action after the spontaneous symmetry breaking will possess a residual $U(1) \times U(1)$ gauge symmetry, with
\begin{equation}
\hat{R} = \begin{bmatrix}
U(1) & 0 & 0\\
0 & U(1) & 0\\
0 & 0 & u
\end{bmatrix}\;,
\end{equation}
where $u$ is a parameter that ensures the matrix has a zero trace.

From the rearrangement of bosonic degrees of freedom promoted by Higgs mechanism, the gauge fields that gain mass after SSB are
\begin{eqnarray}\label{v^2A}
    v^2 A_{\mu}^{aD} A_{\mu}^{aD} =  v^2 A_{\mu}^{a8} A_{\mu}^{a8} = v^2\left(A^1_\mu A^1_\mu+A^2_\mu A^2_\mu+ \cdots + A^7_\mu A^7_\mu\right)\;, 
\end{eqnarray}
see Eq. \eqref{phiDDphi}, while the field $A_\mu^8 A_\mu^8$ remains massless. The direction of $\bar{\eta}^a$ in the vacuum does not affect the Higgs mechanism due to algebra of the operators $(\hat{\bar{\kappa}}, \hat{\kappa}, \hat{\bar{\rho}}, \hat{\rho}$), see Eq. \eqref{algebra}. In other words, the energy scale $v^2$ appears as the only contribution to the mass of the vector gauge bosons after SSB. Due to the scalar supersymmetry of Witten-Fujikawa model, this rearrangement is transferred to the fermionic degrees of freedom, from which some components of the fields $\psi_\mu^a$, $\bar{\eta}^a$ and $\chi^a_{\mu\nu}$ will also display propagators with mass poles, defined  by the energy scale ($v^2$) introduced by the Fujikawa potential.

In the final WF Lagrangean, see equations \eqref{Sssb} (or Eq. \eqref{LBF} for $S_F$ after SSB) and \eqref{SWafterSSB}, the new quadratic terms involving the fermionic fields after SSB are
\begin{eqnarray}\label{quadfermi}
    \mathcal{L}_{\text{quad}}[\psi_\mu, \bar{\eta}, \bar{\chi}_{\mu\nu}] &=& \frac{v_\kappa}{8} f^{Dbc} \bar{\chi}^b_{\mu\nu}\bar{\chi}^c_{\mu\nu}  +\frac{\bar{v}_\kappa}{2}f^{Dbc} \psi^b_\mu \psi^c_\mu + \frac{v_\kappa}{2} f^{Dbc} \bar{\eta}^b\bar{\eta}^c \nonumber\\
&+& \frac{v_{\rho \, 1}}{8} f^{(D-2)bc} \bar{\chi}^b_{\mu\nu}\bar{\chi}^c_{\mu\nu}  +\frac{\bar{v}_{\rho \,1}}{2}f^{(D-2)bc} \psi^b_\mu \psi^c_\mu + \frac{v_{\rho \, 1}}{2} f^{(D-2)bc} \bar{\eta}^b\bar{\eta}^c \nonumber\\
&+& \frac{v_{\rho \, 2}}{8} f^{(D-1)bc} \bar{\chi}^b_{\mu\nu}\bar{\chi}^c_{\mu\nu}  +\frac{\bar{v}_{\rho \,2}}{2}f^{(D-1)bc} \psi^b_\mu \psi^c_\mu + \frac{v_{\rho \, 2}}{2} f^{(D-1)bc} \bar{\eta}^b\bar{\eta}^c \;,
\end{eqnarray}
which suggests the generation of mass to the fields $\psi^a_\mu$, $\bar{\chi}^a$ and $\bar{\eta}^a$. These quadratic terms are not the usual ones given by $\text{Tr}\, \{m \bar{\Psi} \Psi$\}, being $\Psi$ a generic fermionic field with mass $m$. Due to their statistics, the term above represents a gauge-invariant possibility to build quadratic fermionic terms with these fields, with their components contracted with the structure constants, taking one of its colors as a fixed quantum number, and with ($v_\kappa$, $v_{\rho 1}$, $v_{\rho 2}$)  and ($\bar{v}_\kappa$, $\bar{v}_{\rho 1}$, $\bar{v}_{\rho 2}$) compensating the ghost numbers of the fields.  

We are left with the task of demonstrating that the quadratic terms of Eq. \eqref{quadfermi} indeed generate mass poles in the fermionic propagators $\langle \psi^a_\mu \psi^b_\nu \rangle$, $\langle \bar{\chi}^a_{\mu\nu} \bar{\chi}^b_{\lambda\sigma} \rangle$ and $\langle \bar{\eta}^a \bar{\eta}^b \rangle$. As we will see, this is exactly the case.  This result was expected since the bosonic and fermionic degrees of freedom are interconnected. As mentioned in Section 2, the propagating modes of $A_\mu$ have helicities $(1, -1)$, and of $(\phi, \bar{\phi})$, $(0,0)$; while of the fermionic fields $(\bar{\eta}, \psi, \bar{\chi})$, helicities $(1, -1, 0, 0)$. Precisely, it reveals that the bosonic and fermionic degrees of freedom are equal, and this relation is consistently described by the fermionic supersymmetry \eqref{Wittensym}, that defines a one-to-one map between these fields. In practice, the Higgs mechanism establishes a rearrangement of the degrees of freedom. Due to the symmetry between bosons and fermions, such a rearrangement is automatically transferred to the fermionic fields, which also acquires mass, being this phenomenon interpreted as an extension of Higgs mechanism in the supersymmetric topological case.

This way, we are able to compute the mass poles of the fermionic propagators after SSB. The quadratic terms $\bar{\eta} D_\mu \psi$ and $D_\mu \psi_\nu \cdot \bar{\chi}_{\mu\nu}$, present in the original Witten action, perform a mixture between the fermionic fields $\psi_\mu$, $\bar{\eta}$ and $\bar{\chi}_{\mu\nu}$. First, we will consider the propagators considering this mixture in the presence of the quadratic term
\begin{eqnarray}\label{mix1}
\frac{v_\kappa}{8} f^{Dbc} \bar{\chi}^b_{\mu\nu}\bar{\chi}^c_{\mu\nu}  +\frac{\bar{v}_\kappa}{2}f^{Dbc} \psi^b_\mu \psi^c_\mu + \frac{v_\kappa}{2} f^{Dbc} \bar{\eta}^b\bar{\eta}^c\;, 
\end{eqnarray}
that appears after the SSB, see the first line of Eq. \eqref{quadfermi}. (These quadratic terms will be responsible for generating terms of the fermionic propagators in the sector with massive poles.)  It promotes a mixture for $\{b,c\} \neq D$. With $D=8$, it corresponds to the components $\{b,c\} = \{1, \cdots, 7\}$, according to the fluctuations around the nontrivial vacuum with direction $\delta^{a8}$. To correctly mix the components involved in this case, we must assume the Witten quadratic terms $\bar{\eta}^{b} D_\mu^{bc} \psi^c$ and  $D_\mu^{bc} \psi_\nu^c \cdot \bar{\chi}^c_{\mu\nu}$
with $\{b, c\} \neq 8$. This way, the quadratic part of the WF action after SSB, for the fields $\psi_\mu^a$, $\bar{\chi}_{\mu\nu}^a$ and $\bar{\eta}^a$ in this type of mixture are given by 
\begin{eqnarray}
S^{(2)}_{DW}[\psi_\mu, \bar{\chi}_{\mu\nu},\bar{\eta}] &=& \frac{1}{2} \int d^4 x \left( -i \bar{\eta}^a \partial_\mu \psi^a_\mu+i\partial_\mu \psi^a_\nu \bar{\chi}^a_{\mu\nu} + \frac{v_k}{8} f^{Dbc} \bar{\chi}^b_{\mu\nu}\bar{\chi}^c_{\mu\nu} + \frac{\bar{v}_k}{2} f^{Dbc} \psi^b_{\mu} \psi^c_{\mu}\right.\nonumber\\
&+& \left.\frac{v_k}{2} f^{Dbc} \bar{\eta}^b \bar{\eta}^c \right)\;, 
\end{eqnarray}
which, in the Fourier space, yields
\begin{eqnarray}
S^{(2)}_{DW}[\bar{\eta}, \bar{\chi}_{\mu\nu},\psi_\mu] &=& \frac{1}{2} \int d^4 k \left[\frac{1}{2}\bar{\eta}^a(-k) k_\mu \psi^a_\mu(k) - \frac{1}{2} k_\mu  \psi^a_\mu(-k) \bar{\eta}^a(k) - \frac{1}{8} \psi^a_\nu (\varepsilon_{\alpha\beta\nu\mu} k_\mu+\delta_{\nu\alpha} k_\beta - \delta_{\nu\beta} k_\alpha)\bar{\chi}^a_{\alpha \beta}\right.\nonumber\\
&+& \left.\frac{1}{8}  \bar{\chi}^a_{\alpha\beta} (\varepsilon_{\alpha\beta\nu\mu}k_\mu + \delta_{\nu\alpha} k_\beta - \delta_{\nu\beta} k_\alpha)\psi^a_\nu + \frac{v_k}{32} f^{Dbc} \bar{\chi}^a_{\mu\nu}(\delta_{\mu\alpha} \delta_{\nu\beta}-\delta_{\mu\beta} \delta_{\nu\alpha} ) \bar{\chi}^a_{\alpha \beta}\right. \nonumber\\
&+&  \left.\frac{\bar{v}_k}{2} f^{Dbc} \psi^b_{\mu} \psi^c_{\mu}+ \frac{v_k}{2} f^{Dbc} \bar{\eta}^b \bar{\eta}^c   \right] \;.
\end{eqnarray}
Rewriting the equation above in a matrix way, according to 
\begin{eqnarray}
S^{(2)}_{DW}[\bar{\eta}, \bar{\chi}_{\mu\nu},\psi_\mu]&\equiv & \frac{1}{2} \int d^4 x \begin{pmatrix}
\bar{\eta}^a & \psi_\mu^b & \bar{\chi}^c_{\nu \beta} 
\end{pmatrix}
\bigtriangleup^{abc \cdot fgh}_{\mu\nu\beta \cdot \alpha \lambda \sigma} 
 \begin{pmatrix}
\bar{\eta}^f \\
 \psi_\alpha^g \\
  \bar{\chi}^h_{\lambda \sigma} 
\end{pmatrix} \nonumber \\
&\equiv & \frac{1}{2} \int d^4 x \Phi^T \bigtriangleup \Phi\;, 
\end{eqnarray}
one has
\begin{equation}\label{waveoperators}
\bigtriangleup =  \begin{pmatrix}
\frac{v_k}{2} f^{Daf} & \frac{1}{2} k_\alpha \delta^{ag} & 0 \\
-\frac{1}{2} k_\mu \delta^{bf} & \frac{\bar{v}_k}{2} f^{Dbg} \delta_{\mu\alpha} & - \frac{1}{8} \delta^{bh}(\varepsilon_{\lambda\sigma\mu\delta}k_\delta +\delta_{\mu\lambda} k_\sigma - \delta_{\mu \sigma} k_\lambda ) \\
0 & \frac{1}{8} \delta^{cg} (\varepsilon_{\nu\beta\alpha\delta}k_\delta + \delta_{\nu\alpha}k_\beta - \delta_{\beta \alpha} k_\nu ) & \frac{v_k}{16} f^{Dch} (\delta_{\nu\lambda} \delta_{\beta \sigma} - \delta_{\nu\sigma} \delta_{\beta \lambda}) 
\end{pmatrix}\;.
\end{equation}
The  construction of the matrix \eqref{waveoperators} for the fermionic wave operators is valid for $\text{color} =  \{1,\cdots,7\}$. The propagators of the Abelian fields $\psi^8_\mu$, $\chi^8_{\mu\nu}$ and $\eta^8$ do not appear here, and have no chance of acquiring mass via Higgs mechanism in this case. From the wave operator matrix \eqref{waveoperators}, we infer that the matrix for the propagators takes the form
\begin{equation} \label{propagators}
\bigtriangleup^{-1} = \begin{pmatrix}
A^{f a^\prime} & C^{fb^\prime}_{\mu^\prime} & 0 \\
B^{g a^\prime}_\alpha & D^{gb^\prime}_{\mu^\prime \alpha} & F^{gc^\prime}_{\nu^\prime \beta^\prime \alpha} \\
0 & E^{hb^\prime}_{\mu^\prime \lambda \sigma} & G^{h c^\prime}_{\nu^\prime \beta^\prime \lambda \sigma} \\
\end{pmatrix}\;, 
\end{equation}
such that
\begin{equation} \label{inverse}
\bigtriangleup \bigtriangleup^{-1} = \begin{pmatrix}
\delta^{a a^\prime} & 0 & 0 \\
0 & \delta^{bb^\prime}\delta_{\mu\mu^\prime} & 0 \\
0 & 0 & \frac{1}{2}\delta^{cc^\prime} (\delta_{\nu\nu^\prime}\delta_{\beta\beta^\prime} - \delta_{\nu\beta^\prime}\delta_{\beta\nu^\prime}) 
\end{pmatrix}\;.
\end{equation}

Accordingly to the symmetry of color and Lorentz indices, including the self-dual property of $\bar{\chi}_{\mu\nu}$, the most general expressions for the propagators given by the elements of matrix \eqref{propagators}, are given by
\begin{eqnarray}
A^{f a^\prime} &=& a f^{Dfa^\prime}\;,\label{A} \\
B^{g a^\prime}_\alpha & =&  k_\alpha (b_1\delta^{g a^\prime}+b_2f^{Dga^\prime})\;,\label{B2} \\
C^{f b^\prime}_{\mu^\prime} & =&  - B^{f b^\prime}_{\mu^\prime}, \label{C}\\
D^{g b^\prime}_{\mu^\prime\alpha} & =&  f^{Dgb^\prime} (d_1\mathbb{T}_{\mu^\prime\alpha}+d_2\mathbb{L}_{\mu^\prime\alpha})\;,\label{D} \\
E^{hb^\prime}_{\mu^\prime \lambda \sigma} &=& \delta^{hb^\prime}\left[ e_1 \varepsilon_{\mu^\prime \lambda \sigma \delta} k_\delta + e_2\left( \delta_{\mu^\prime \lambda} k_\sigma - \delta_{\mu^\prime \sigma} k_\lambda\right) \right]+\nonumber\\
&+& f^{Dhb^\prime}\left[ e_3 \varepsilon_{\mu^\prime \lambda \sigma \delta} k_\delta + e_4\left( \delta_{\mu^\prime \lambda} k_\sigma - \delta_{\mu^\prime \sigma} k_\lambda\right) \right]\;,\label{E}\\
F^{gc^\prime}_{\nu^\prime \beta^\prime \alpha} &=&-E^{gc^\prime}_{\nu^\prime \beta^\prime \alpha}\;,\label{F}\\
G^{hc^\prime}_{\nu^\prime \beta^\prime \lambda \sigma} &=& \delta^{hc^\prime} \left[ g_1 \varepsilon_{\nu^\prime \beta^\prime \lambda \sigma} + g_2 \left(\delta_{\nu^\prime \lambda} \delta_{\beta^\prime \sigma} - \delta_{\nu^\prime \sigma} \delta_{\beta^\prime \lambda}\right) + h_1 \left(\varepsilon_{\lambda \sigma \nu^\prime \delta} k_\delta k_{\beta^\prime} - \varepsilon_{\lambda \sigma \beta^\prime \delta} k_\delta k_{\nu^\prime} \right)+\right.\nonumber\\
&+&\left. h_2\left( \varepsilon_{\nu^\prime \beta^\prime \lambda \delta} k_\delta k_{\sigma} - \varepsilon_{\nu^\prime \beta^\prime \sigma \delta} k_\delta k_{\lambda}\right)+h_3 \left( \delta_{\lambda \beta^\prime} k_\sigma k_{\nu^\prime}- \delta_{\sigma \beta^\prime} k_\lambda k_{\nu^\prime} + \delta_{\sigma \nu} k_\lambda k_{\beta^\prime} - \delta_{\lambda \nu^\prime} k_\sigma k_{\beta^\prime} \right)\right]+ \nonumber\\
&+& f^{Dhc^\prime} \left[ g_3 \varepsilon_{\nu^\prime \beta^\prime \lambda \sigma} + g_4 \left(\delta_{\nu^\prime \lambda} \delta_{\beta^\prime \sigma} - \delta_{\nu^\prime \sigma} \delta_{\beta^\prime \lambda}\right)+ h_4 \left(\varepsilon_{\lambda \sigma \nu^\prime \delta} k_\delta k_{\beta^\prime} - \varepsilon_{\lambda \sigma \beta^\prime \delta} k_\delta k_{\nu^\prime} \right)+\right.\nonumber\\
&+&\left. h_5\left( \varepsilon_{\nu^\prime \beta^\prime \lambda \delta} k_\delta k_{\sigma} - \varepsilon_{\nu^\prime \beta^\prime \sigma \delta} k_\delta k_{\lambda}\right)+h_6 \left( \delta_{\lambda \beta^\prime} k_\sigma k_{\nu^\prime}- \delta_{\sigma \beta^\prime} k_\lambda k_{\nu^\prime} + \delta_{\sigma \nu} k_\lambda k_{\beta^\prime} - \delta_{\lambda \nu^\prime} k_\sigma k_{\beta^\prime} \right) \right]\;,\nonumber\\
\label{G}
\end{eqnarray}
where the coefficients are real parameters, with $\mathbb{L}_{\mu\nu} = \frac{k_\mu k_\nu}{k^2}$ and $\mathbb{T}_{\mu\nu} = \delta_{\mu\nu} - \frac{k_\mu k_\nu}{k^2}$. The propagator  $G^{ab}_{\mu\nu\lambda\sigma} = \langle \chi^a_{\mu\nu}\chi^b_{\lambda\sigma} \rangle$, under the permutation $[\mu \nu]\Leftrightarrow [\lambda \sigma]$ and $ a \Leftrightarrow b$, transforms as  
\begin{equation}
 \langle \chi^a_{\mu\nu}\chi^b_{\lambda\sigma} \rangle \rightarrow -  \langle \chi^b_{\lambda\sigma}\chi^a_{\mu\nu} \rangle\;, \label{symmetry}
\end{equation} 
which implies on
\begin{eqnarray}
g_1 = g_2 = h_3  &=& 0\;,\label{symchi1}\\
h_2 &=& -h_1\;,\\
h_5 &=& h_4\;.\label{symchi3}
\end{eqnarray}

The elements of the matrix \eqref{propagators}, namely, $A^{a b}$, $B^{ab}_\mu$, $D^{ab}_{\mu\nu}$ $E^{ab}_{\mu\lambda\sigma}$ and $G^{ab}_{\nu\beta\lambda\sigma}$ represent, respectively, the propagators $\langle \bar{\eta}^a\bar{\eta}^b \rangle$, $\langle \bar{\eta}^a \psi^b_\mu\rangle$, $\langle \psi^a_\mu \psi^b_\nu \rangle$, $\langle \psi^a_\mu \bar{\chi}^b_{\lambda\sigma} \rangle$ and $\langle \bar{\chi}^a_{\mu\nu} \bar{\chi}^b_{\lambda\sigma} \rangle$. Replacing \eqref{waveoperators} and \eqref{propagators} into \eqref{inverse}, one gets the following system of equations :
\begin{eqnarray}
\frac{v_k}{2} f^{Daf} A^{fa^\prime} + \frac{1}{2} k_\alpha \delta^{ag} B^{ga^\prime}_\alpha &=&  \delta^{aa^\prime} \;, \label{eq1a} \\
\frac{v_k}{2} f^{Daf} C^{fb^\prime}_{\mu^\prime} + \frac{1}{2} k_\alpha \delta^{ag} D^{gb^\prime}_{\mu^\prime \alpha} &=&  0\;,\label{eq1b}  \\
\frac{1}{2} k_\alpha \delta^{ag} F^{gc^\prime}_{\nu^\prime \beta \alpha} &=& 0\;,\label{eq1c}\\ 
-\frac{1}{2} k_\mu \delta^{bf} A^{fa^\prime} + \frac{\bar{v}_k}{2} f^{Dbg} \delta_{\mu\alpha} B^{ga^\prime}_\alpha &=&  0 \;,\label{eq2a} \\
-\frac{1}{2} k_\mu \delta^{bf} C^{fb^\prime}_{\mu^\prime} +\frac{\bar{v}_k}{2} f^{Dbg} \delta_{\mu\alpha} D^{gb^\prime}_{\mu^\prime \alpha} - \frac{1}{8} \delta^{bh} (\varepsilon_{\lambda\sigma\mu\delta}k_\delta+\delta_{\mu\lambda} k_\sigma - \delta_{\mu \sigma} k_\lambda ) E^{h b^\prime}_{\mu^\prime \lambda \sigma} &=&  \delta^{bb^\prime} \delta_{\mu\mu^\prime}\;, \label{eq2b} \\
\frac{\bar{v}_k}{2} f^{Dbg} \delta_{\mu\alpha}  F^{gc^\prime}_{\nu^\prime \beta^\prime \alpha} - \frac{1}{8} \delta^{bh} (\varepsilon_{\lambda\sigma\mu\delta}k_\delta+\delta_{\mu\lambda} k_\sigma - \delta_{\mu \sigma} k_\lambda ) G^{hc^\prime}_{\nu^\prime \beta^\prime\lambda \sigma} &=& 0\;,  \label{eq2c}\\
\frac{1}{8} \delta^{cg} (\varepsilon_{\nu\beta\alpha\delta}k_\delta+\delta_{\nu\alpha} k_\beta - \delta_{\beta \alpha} k_\nu) B^{ga^\prime}_\alpha &=&  0 \;,\label{eq3a} \\
\frac{1}{8} \delta^{cg} (\varepsilon_{\nu\beta\alpha\delta}k_\delta+\delta_{\nu\alpha} k_\beta - \delta_{\beta \alpha} k_\nu) D^{gb^\prime}_{\mu^\prime \alpha} + \frac{v_k}{16} f^{Dch} (\delta_{\nu\lambda} \delta_{\beta\sigma} - \delta_{\nu \sigma} \delta_{\beta \lambda} ) E^{h b^\prime}_{\mu^\prime \lambda \sigma} &=&  0\;,\label{eq3b}  \\
\frac{1}{8} \delta^{cg} (\varepsilon_{\nu\beta\alpha\delta}k_\delta+\delta_{\nu\alpha} k_\beta - \delta_{\beta \alpha} k_\nu) F^{gc^\prime}_{\nu^\prime \beta^\prime \alpha} + \frac{v_k}{16}f^{Dch} (\delta_{\nu\lambda} \delta_{\beta\sigma} - \delta_{\nu \sigma} \delta_{\beta \lambda} ) G^{hc^\prime}_{\nu^\prime \beta^\prime\lambda \sigma}
&=& \frac{1}{2}\delta^{cc^\prime} (\delta_{\nu\nu^\prime}\delta_{\beta\beta^\prime} - \delta_{\nu\beta^\prime}\delta_{\beta\nu^\prime})  \;.\nonumber\\
\label{eq3c}
\end{eqnarray}

Using equations \eqref{A}-\eqref{G}, together with  \eqref{symchi1}-\eqref{symchi3}, the system for the coefficients can be easily solved and yields
the following terms of the fermionic propagators in the sector with massive poles:
\begin{eqnarray}
\langle \bar{\eta}^a \bar{\eta}^b \rangle(k) &=& f^{Dab} \frac{2 \bar{v}_k}{k^2-v^2}\;,\label{fp1} \\
\langle \bar{\eta}^a \psi_\mu^b \rangle(k) & =&  \delta^{ab}  k_\mu \frac{2}{k^2-v^2}\;,\label{B} \\
\langle \psi_\mu^a  \psi^b_\nu \rangle(k)  & =&  f^{Dab}\frac{2 v_k}{k^2-v^2} (\mathbb{T}_{\mu \nu}+\mathbb{L}_{\mu\nu})\;, \\
\langle \psi_\mu^a \bar{\chi}^b_{\lambda \sigma} \rangle(k) &=& -2\delta^{ab}\left[ \frac{1}{k^2-v^2}\varepsilon_{\mu \lambda \sigma \delta} k_\delta + \frac{1}{k^2-v^2}\left( \delta_{\mu \lambda} k_\sigma - \delta_{\mu \sigma}k_\lambda \right)\right] \;,\\
\langle \bar{\chi}^a_{\nu \beta} \bar{\chi}^b_{\lambda \sigma}\rangle(k) &=& -f^{Dab}\frac{2}{v^2} \left[  \bar{v}_k\left( 1- \frac{v^2}{k^2-v^2}\right) \left(\delta_{\nu \lambda} \delta_{\beta \sigma} - \delta_{\nu \sigma} \delta_{\beta \lambda}\right)+\right.\nonumber\\
&-&\left.  \frac{ \bar{v}_k}{k^2-v^2} \left(\varepsilon_{\lambda \sigma \nu \delta} k_\delta k_{\beta} - \varepsilon_{\lambda \sigma \beta \delta} k_\delta k_{\nu} +  \varepsilon_{\nu \beta \lambda \delta} k_\delta k_{\sigma} - \varepsilon_{\nu \beta \sigma \delta} k_\delta k_{\lambda}\right) \right]\;,\label{fp5}
\end{eqnarray}
with $a,b = \{1, \cdots, 7\}$ in this case. 

This result shows the mass generation for the fermionic fields $\psi^a_\mu$, $\bar{\chi}^a_{\mu\nu}$ and $\bar{\eta}^a$ with $a \neq 8$. As discussed in the previous section, this was the expected result for the Higgs mechanism acting together the fermionic scalar supersymmetry \eqref{Wittensym}. One must note that the mass poles for the fermionic fields are also described in terms of the energy scale introduced by the potential \eqref{Ussb} that is precisely the one that appears as the mass of the vector gauge boson, see \eqref{v^2A}. If we have considered other mixtures involving the component $a=8$,  taking into account the second and third line of \eqref{quadfermi}, there would be no contribution to the mass poles. The fermionic poles with mass $m^2_F = v^2$ appear as a consequence of the product $\bar{v}_\kappa v_\kappa = v^2$. Taking into account the other mixtures, the results \eqref{fp1}-\eqref{fp5} would receive contributions proportional to $(\bar{v}_{\rho 1}, v_{\rho 1})$ or $(\bar{v}_{\rho 2}, v_{\rho 2})$, which would have produced poles with mass $m^2= \bar{v}_{\rho 1} v_{\rho 1}$ or $\bar{v}_{\rho 2}, v_{\rho 2}$, which are zero, due to the fermionic nature of $(H_1, H_2)$, $H_1^2 = H_2^2 = 0$. The results \eqref{fp1}-\eqref{fp5}  could be seen as the final expression for states where $(v_\kappa, \bar{v}_\kappa) \gg (v_{\rho i},\bar{v}_{\rho i} )$. In any case, the mixture \eqref{mix1} is the only capable of generating terms in the propagators with massive poles for the fermions $(\psi^a_\mu, \bar{\chi}^a_{\mu\nu}, \bar{\eta}^a)$. From the analysis of this sector, we can conclude that the propagators with components $(\psi^8_\mu, \bar{\chi}^8_{\mu\nu}, \bar{\eta}^8)$ do not gain mass after the SSB, in the same way as the boson $A^8_\mu$. The other components are interconnected by supersymmetry, which correlates the fermionic and bosonic degrees of freedom, which are consistently rearranged by Higgs mechanism. The appearance of massless fermions  $(\psi^8_\mu, \bar{\chi}^8_{\mu\nu}, \bar{\eta}^8)$ can be understood as a consequence of Goldstone theorem.

\section{Gauge symmetries, scalar supersymmetry, and broken topological phase}

In a more generic way, to obtain a vacuum that minimizes the Fujikawa's potential, we could consider the vacuum directions
\begin{equation}\label{na2}
v^a = \begin{pmatrix}
v_1 \\
v_2 \\
\vdots\\
v_D\\
\end{pmatrix}\;, \quad \bar{\eta}^a  = \begin{pmatrix}
H_1 \\
H_2 \\
\vdots\\
H_D
\end{pmatrix}\;,
\end{equation}
such that the components of $\phi^a$ and $\bar{\phi}^a$ for $a = \{1, \cdots, D\}$, being the $D$ the dimension of the gauge group $G$,  would oscillate around the nontrivial vacuum. (Depending on the number of generators in the subalgebra of Cartan, one must set some vacuum directions to zero.) To break the scalar supersymmetry, one needs, at least, one non-zero direction of $v^a$, and two non-zero and distinct directions of $\bar{\eta}^a$, according to the $\delta$-transformation of $\phi_0$ and $\bar{\phi}_0$ defined by equations Eq. \eqref{deltaphi01} and \eqref{deltaphi02}. In the previous section, we considered the simplest possible case. As the vacuum solution must belong to the subalgebra, the number of generators  that commute between each other defines the gauge symmetries that could break the fermionic supersymmetry. Assuming that $G = SU(N)$, it could only be broken if $N \geq 3$, since the $SU(3)$ group possesses 3 generators in the Cartan subalgebra. The values of the parameters $v_i$ and $H_i$ can also influence the type of SSB. In the $SU(3)$ case for $v \neq H_1 \neq H_2$, the symmetry breaking occurs from $SU(3)$ to a final theory with gauge symmetry  $U(1) \times U(1)$. This breaking is not general, and depends on the relation between the nontrivial vacuum directions.    

For the general directions above, we could write the components of the bosonic scalar fields in the generic form 
\begin{eqnarray}\label{phinagen}
\phi^a(x)\vert_{i-component} &=& (\omega_i + v_{0\, i} )\delta^{ai}\;, \label{phinagen1} \\
\bar{\phi}^a(x)\vert_{i-component}  &=& (\bar{\omega}_i + \bar{v}_{0 \, i})\delta^{ai}\;,\label{phinagen2}
\end{eqnarray}
where
\begin{eqnarray}
    v_{0\, i} &=& \kappa v_i + \rho H_i\\
    \bar{v}_{0\, i} &=& \bar{\kappa} v_i + \bar{\rho} H_i\;,
\end{eqnarray}
with $i = \{1, \cdots, D\} $,  being $(\omega_i, \bar{\omega}_i)$ the vacuum fluctuations of the corresponding $i$-components of $(\phi, \bar{\phi})$. 

After the SSB, it will occur the mass generation for gauge bosons and fermions, with the rearrangement of the degrees of freedom being  explained by the Higgs mechanism in the presence of the scalar supersymmetry. The mechanism works perfectly as in the ordinary YM for each direction with nontrivial vacuum due to the condition \eqref{cond}, which states that $\bar{v}_{0\, i}v_{0\,i} = v^2$, which appears as the mass of $A_\mu^2$, with the non-zero components of $v^a$ in Cartan's subalgebra defining the components of $A_\mu^a$ that will remain massless, while the others will gain mass. It demonstrates that the Higgs mechanism in the topological case is directly related to the algebra of the operators $(\hat{\kappa}, \hat{\bar{\kappa}},\hat{\rho}, \hat{\bar{\rho}})$ given by Eq. \eqref{algebra}.  

The scalar supersymmetry connects the fermionic and bosonic degrees of freedom. The SSB promotes the rearrangement of these degrees, that also involves the appearance of fermionic quadratic terms, with the number of fermionic fields affected by the SSB energy scale given by the same number of massive vector gauge bosons. Writing the fluctuations in the form $\omega =  \omega_i \delta^{a i} \frac{T^a}{2}$, $\bar{\omega} =  \bar{\omega}_i \delta^{a i} \frac{T^a}{2}$,  around the nontrivial vacuum $v_0 = v_{0\,i} \delta^{a i} \frac{T^a}{2}$, $\bar{v}_0 = \bar{v}_{0\, i} \delta^{a i} \frac{T^a}{2}$, the Witten action after the SSB in curved spacetime reads
\begin{eqnarray}\label{SWitten2}
S_{W}\vert_{\text{after SSB}}&=& \frac{1}{g_0^2}Tr\int d^4x \sqrt{-g} \{\frac{1}{4} F_{\mu\nu}^{+}F^{\mu\nu\,+} + \frac{1}{2} \bar{\omega}D_{\mu} D^{\mu}\omega -i\bar{\eta}D_{\mu} \psi^\mu + i D_{\mu} \psi_\nu \cdot \bar{\chi}^{\mu\nu}  \nonumber\\
&-& \frac{i}{8} \omega \left[ \bar{\chi}_{\mu\nu}, \bar{\chi}^{\mu\nu}\right]-\frac{i}{2} \bar{\omega} \left[ \psi_\mu, \psi^\mu \right] - \frac{i}{2} \omega \left[ \bar{\eta}, \bar{\eta}\right] - \frac{1}{8} \left[ \omega, \bar{\omega}\right]^2 + \mathcal{L}^{(W)}_{B}\}\;,\nonumber\\
&=& S_W[\omega, \bar{\omega}, \text{other}] + \frac{1}{g_0^2}Tr\int d^4x \sqrt{-g}\mathcal{L}^{(W)}_B\;. 
\end{eqnarray}
where $S_W[\omega, \bar{\omega}, \text{other}]$ is the original Witten action \eqref{SWitten}, just replacing $\phi$ and $\bar{\phi}$ by the fluctuations $\omega$ and $\bar{\omega}$ for the bosonic fields with nontrivial vacuum, respectively, and $\mathcal{L}_B$, given by 
\begin{eqnarray}\label{LB}
\mathcal{L}^{(W)}_{B} &=& \text{Tr} \left\{\frac{1}{2} \bar{\omega}D_{\mu} D^{\mu} v_0 + \frac{1}{2} \bar{v}_0 D_{\mu} D^{\mu} v_0 + \frac{1}{2} \bar{v}_0 D_{\mu} D^{\mu} \omega- \frac{i}{8} v_0 \left[ \bar{\chi}_{\mu\nu}, \bar{\chi}^{\mu\nu}\right]-\frac{i}{2} \bar{v}_0 [\psi_\mu, \psi^\mu]\right. \nonumber\\
&-& \left. \frac{i}{2} v_0 [\bar{\eta}, \bar{\eta}]+\frac{1}{2}\left[\left([\bar{\omega}, \omega] +[\bar{\omega}, v_0] +  [\bar{v}_0,\omega]+ [\bar{v}_0,v_0 ]\right)^2 - \left([\bar{\omega}, \omega]\right)^2\right]\right\}\,.
\end{eqnarray}
For consistency, the fluctuations $(\bar{\omega}, \omega)$ transform in the same way as the fields $(\bar{\phi}, \phi)$, which are the fluctuations in the limit $(\kappa, \bar{\kappa}, \rho, \bar{\rho}) \rightarrow (0,0,0,0)$, while the $\delta$-transformation of the vacuum in the directions \eqref{na2} yields, see equations \eqref{deltaphi01} and \eqref{deltaphi02},
\begin{eqnarray}\label{deltav0}
    \delta v_0 =  \frac{1}{2}\varepsilon\alpha\kappa  [\Upsilon,\xi]\;, \quad     \delta \bar{v}_0 = \frac{1}{2}\varepsilon \alpha \bar{\kappa}  [\Upsilon,\xi]\;,  
\end{eqnarray}
where it is implicit that $[\Upsilon, \xi] = [\Upsilon, \xi]^i \frac{{T}^i}{2}$ with $[\Upsilon, \xi]^i = f^{ijk}\Upsilon^j \xi^k$. Applying Eq. \eqref{deltav0} to Eq. \eqref{LB}, it is easy to check that $\delta \mathcal{L}^{(W)}_{B} \neq 0 $. This result is immediate since the vacuum, that does not possess a symmetry of the original action, breaks precisely this symmetry in the resulting action after the SSB, and the vacuum given by the directions \eqref{na2} is not invariant under the $\delta$-transformation. 

Performing the same analysis for the Fujikawa's action, after replacing the components of $\phi^a$ and $\bar{\phi}^a$ with nontrivial vacuum using \eqref{phinagen1} and \eqref{phinagen2}, one finds
\begin{eqnarray} \label{SFafterSSB}
    S_F\vert_{\text{after SSB}} = S_F[\bar{\omega},\omega, \text{other}] + \frac{1}{g^2_0}\int d^4x \sqrt{-g} \mathcal{L}^{(F)}_B \;,
\end{eqnarray}
with
\begin{eqnarray}\label{LBF}
   \mathcal{L}^{(F)}_B &=& \text{Tr} \{ \xi \cdot \bar{\eta} \left(\bar{\omega} \cdot v_0 + \bar{v}_0 \cdot \omega + v^2 \right) + \Xi \cdot \left([\omega,\bar{v}_0]+ [v_0,\bar{\omega}] + [v_0, \bar{v}_0]\right)\left(\bar{\omega} \cdot v_0 + \bar{v}_0 \cdot \omega + v^2 \right)\nonumber\\
   &-&  2i\left(\Xi \cdot \bar{\eta})(\bar{\eta} \cdot v_0\right) - v^2 \Xi \cdot \left([\omega,\bar{v}_0]+ [v_0,\bar{\omega}] + [v_0, \bar{v}_0]\right) + (\theta \cdot v_0)(\Upsilon \cdot \Xi)+\frac{1}{2} \theta \cdot [\Upsilon, v_0] \nonumber\\
   &+& (\Theta\cdot v)(\Upsilon \cdot \xi) + (\zeta \cdot \bar{v}_0)(\Upsilon \cdot \Xi) - \frac{1}{2} \zeta \cdot [\Upsilon, \bar{v}_0] + (Z \cdot \bar{v}_0)\cdot (\upsilon \cdot \xi)\}\;, 
\end{eqnarray}
so that, applying \eqref{deltav0} to \eqref{LBF}, one concludes that $\delta_T  \mathcal{L}^{(F)}_B \neq 0$ and $\delta_T  \mathcal{L}^{(F)}_B \neq - \delta_T  \mathcal{L}^{(W)}_B$. This way, the $\delta_T$-symmetry of the total action is broken by the vacuum that is not invariant under $\delta$-transformations, as expected in the SSB process.  

Naturally, the break of the fermionic supersymmetry leads to the release of the local degrees of freedom, that were ``confined" in the topological phase. Before the SSB, the only observables were the global ones, given by topological invariants, which were protected by the equivariant cohomology of the $\delta$-symmetry, with the energy-momentum tensor given by a $\delta$-exact term. Precisely, before the SSB one has 
\begin{eqnarray}\label{Tmunuzero}
T_{\mu\nu}^{(0)} = T_{\mu\nu}^{(W)} + T_{\mu\nu}^{(F)}\;, 
\end{eqnarray}
where $T_{\mu\nu}^{(W)}$ is the $\delta$-exact contribution for the energy-momentum that comes from Witten action, see equations \eqref{deltaTexact} and \eqref{Vmunu}, while $T_{\mu\nu}^{(F)}$ is also $\delta$-exact, given by
\begin{eqnarray}
T_{\mu\nu}^{(F)} = \delta_T V_{\mu\nu}^{(F)}\;,     
\end{eqnarray}
with
\begin{eqnarray}
V_{\mu\nu}^{(F)} = -g_{\mu\nu} \bar{\varepsilon}  \left[ \Xi^a \bar{\eta}^a \left(\bar{\phi}^b \phi^b - v^2\right)  + \Theta^a \Upsilon^b\left( \phi^a \Xi^b + \frac{1}{2} f^{abc}\phi^c \right)+Z^a \Upsilon^b\left( \bar{\phi}^a \Xi^b - \frac{1}{2} f^{abc}\bar{\phi}^c \right)\right]\;,    
\end{eqnarray}
according to Fujikawa's action $S_F$, see Eq. \eqref{Sssb}, assuming an arbitrary orientable Riemannian manifold.  

This subtle topological structure has been broken at its base after the SSB. In the broken topological phase, the energy-momentum tensor is not $\delta$-exact anymore, which can be concluded from the contribution to $T_{\mu\nu}$ that comes from $\mathcal{L}^{(W)}_B$ and $\mathcal{L}^{(F)}_B$. A straightforward calculation shows that  
\begin{equation}    T_{\mu\nu}\vert_{\text{after SSB}} = T_{\mu\nu}^{(0)} + \lambda_{\mu\nu} -g_{\mu\nu} \mathcal{L}_B\; ,
\end{equation}
where $T^{(0)}_{\mu\nu}$ is the energy-momentum tensor before the SSB with $(\bar{v}_0, v_0) = (0,0)$, defined by Eq. \eqref{Tmunuzero}, only replacing $(\bar{\phi}, \phi)$ by $(\bar{\omega}, \omega)$, $\mathcal{L}_B = \mathcal{L}^{(W)}_B + \mathcal{L}^{(F)}_B$, and 
\begin{eqnarray}
\lambda_{\mu\nu} = \text{Tr}\, \{\bar{\omega} D_\mu D_\nu v_0 + \bar{v}_0 D_{\mu}D_{\nu} v_0 + \bar{v}_0  D_\mu D_\nu \omega  - \frac{1}{2} g^{\alpha \beta} \left[\bar{\chi}_{\mu\alpha}, \bar{\chi}_{\nu \beta}\right] - i \bar{v}_0\left[ \psi_\mu, \psi_\nu\right] \}\;. 
\end{eqnarray}
The term $T^{(0)}_{\mu\nu}$ is $\delta_T$-exact but the others break the $\delta_T$-symmetry of $T_{\mu\nu}$, with $\delta_T T_{\mu\nu}\vert_{\text{after SSB}} \neq 0$,  as it can be easily checked. Thus,
\begin{equation}    T_{\mu\nu}\vert_{\text{after SSB}}  \neq \delta_T (\text{something})\; ,
\end{equation}
and, consequently, 
\begin{eqnarray}\label{deltagSSB}
    \frac{\delta}{\delta g_{\mu\nu}} Z\vert_{\text{after SSB}} \neq 0\;, 
\end{eqnarray}
meaning that the theory is not invariant under variations on the metric tensor after the SSB. Besides that, we will observe the mass generation for the fermionic fields, which can be understood from the point of view of the Higgs mechanism combined with the scalar supersymmetry, according to the algebra \eqref{algebra}.

\section{Final considerations}

This Witten-Fujikawa model is valid in curved spacetimes, since the commutators of covariant derivatives still appear only acting in the scalar field $\phi$ after introducing the Fujikawa potential, which implies that it can be extended to these spaces without destroying the topological properties of original twisted $\mathcal{N}=2$ SYM theory.  However, before SSB, no local physics can be attached to the theory, since it is invariant under variations on the metric tensor $g_{\mu\nu}$, meaning that all observables must be metric independent. (We have not studied the preservation of the Donaldson invariants as the observables of the WF action for $SU(N)$ gauge theories, in the presence of the new BRST doublets $(\Xi,\, \xi)$, $(\bar{\Theta},\, \bar{\theta})$ and ($Z, \zeta$), and the field $\Upsilon$, see Section 3, since it is not relevant to our final result.)  In order to release the local degrees of freedom, without explicit breaking the fermionic scalar supersymmetry \eqref{Wittensym}, we applied Fujikawa's SSB mechanism, which consists of introducing a potential that belongs to the trivial part of the $\delta$-cohomology. 

After introducing the additional Fujikawa term to DW action, given by Eq. \eqref{Sssb}, one obtains a theory that is not exactly the twisted $\mathcal{N}=2$ SYM. We have introduced a $\delta$-exact term after the twist transformation by introducing three BRST doublets. These BRST doublets preserve the equation of motion of the $\bar{\chi}_{\mu\nu}$ field that is required to prove the nilpotence of the charge $\mathcal{Q}$ -- see Section 2; but it has introduced new fields that could influence the global observables in a nontrivial way. These observables are now defined in terms of a total $\delta_T$-cohomology, see Eq. \eqref{deltaT}. Usually, BRST doublets do not affect observables because of the doublet theorem \cite{Dixon:1991wi,Piguet:1995er,Vandersickel:2011zc}, but a $\delta$-exact term could contain  nontrivial topology. In any case, the final WF action before SSB is automatically topological due to the scalar supersymmetry, and by the fact that the energy-momentum tensor is still given by a $\delta$-exact term, see Section 4.1. These two properties ensure that the theory is a TQFT, with the functional generator being invariant under variations in the metric. The observables are certainly topological invariants, with the action $S_W + S_{F}$ representing a topological phase in quantum field theory, this being sufficient for the purpose of our present work.

Precisely, by adding the $\delta$-exact term $S_{F}$, see Eq. \eqref{Sssb}, to Witten's action $S_W$, we were able to construct the potential $U_{SSB}$ in the bosonic sector, see Eq. \eqref{Ussb}. This construction was possible by exploring the $\delta$-transformation of the $\bar{\eta}$ field for two reasons: (i) it introduces a term in the bosonic sector $(\phi, \bar{\phi})$ without mixtures with other fields; (ii) it provides a nontrivial vacuum solution that is not invariant under the $\delta$-supersymmetry. This way, the Fujikawa potential introduces an energy scale ($v^2$) that will affect the bosonic and fermionic propagators via Higgs mechanism. The vacuum solution can be seen in equations \eqref{vacuum1} and \eqref{vacuum2}, that satisfies the system of equations \eqref{minU1}-\eqref{minU3}.  In Section 4, we studied the spontaneous symmetry breaking for fluctuations around this nontrivial vacuum. To also break the supersymmetry, the vacuum solution requires oscillations in at least three different directions. Otherwise, the vacuum will be invariant under $\delta$-transformations. For this reason, assuming a starting theory with gauge symmetry $G = SU(N)$, we must have $N \geq 3$ in the topological case. Choosing the directions given by Eq. \eqref{na}, it will generate $D-1$ massive gauge vector bosons via Higgs mechanism, with mass $m^2_B = v^2$. The oscillations around the direction $H_i$ do not affect the mass poles of $A_\mu$, which is ensured by the algebra of operators ($\hat{\bar{\kappa}}, \hat{\kappa}, \hat{\bar{\rho}}, \hat{\rho}$) defined in Section 3, see Eq. \eqref{algebra}. Such an algebra will act together with Higgs mechanism, leading to a rearrangement of fermionic and bosonic degrees of freedom in a consistent way.  

Due to the scalar supersymmetry, the fermionic and bosonic degrees of freedom are interconnected. The supersymmetry is also broken in this SSB process, see Eq. \eqref{deltav0}, such that it implies the release of local degrees of freedom. After SSB, the theory is no longer metric independent. The energy-momentum tensor is not $\delta$-exact anymore, see Section 5, where we have analyzed the breaking for a more generic vacuum. As a consequence, the Higgs mechanism will also provide the mass generation to the fermionic fields $(\psi_\mu, \bar{\eta}, \bar{\chi}_{\mu\nu})$. In Section 4, we also computed the fermionic propagators after SSB for a particular case assuming a symmetry breaking of the type $SU(3) \rightarrow U(1) \times U(1)$, with the vacuum directions obeying $\overrightarrow{v}\cdot \overrightarrow{H} = 0$. Due to the algebra \eqref{algebra}, the gauge bosons that acquire mass are $A_\mu^a$ with $a = \{1,\cdot,7\}$. We observe the existence of mass poles for the same components of the fermionic fields, with the same mass, $m^2_F = m^2_B= v^2$, see equations \eqref{fp1}-\eqref{fp5}, in accordance with Higgs mechanism in the presence of the scalar supersymmetry. The poles involving the components $(\psi_\mu^8, \bar{\eta}^8, \bar{\chi}_{\mu\nu}^8)$ remain massless, similarly to  $A_\mu^8$. The massive fermionic poles only appear due to the energy scale introduced by Fujikawa potential. Without it, all fermionic propagators in the twisted $\mathcal{N} = 2$ SYM theory are massless, see \cite{Brooks:1988jm}.   

\begin{figure}[!htb]
	\centering
\includegraphics[scale=0.7]{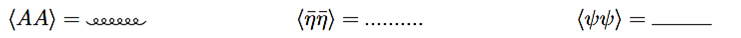}
\caption{Propagators.}
    \label{fig1}
\end{figure}

\begin{figure}[!htb]
	\centering
\includegraphics[scale=0.7]{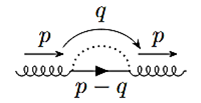}
\caption{Feynman diagram with a fermionic loop.}
    \label{fig2}
\end{figure}

In the twisted $\mathcal{N} = 2$, the fermionic fields are not unitary \cite{Witten:1988ze}. We expect the same in the WF case. This does not rule out the possibility of these fields acting as ghost fields, giving quantum contributions to the gauge field propagator and $\langle A A \cdots A \rangle$ scatterings. Such radiative corrections could influence the $\beta$-function, indicating how a topological phase of matter would affect the running of the gauge coupling and the local dynamics of the system at the quantum level. For instance, defining the propagators as in Fig. \ref{fig1},
we could consider the 1-loop contribution to the gauge propagator $\langle A^a_\mu A^b_\nu \rangle$(p) given by Fig. \ref{fig2}, with the vertex corresponding to the interaction  $g\text{Tr} \left(\bar{\eta} [A_\mu, \psi_\mu]\right)$. One concludes that the gauge field Green functions could suffer quantum corrections from the energy scale $v^2$ present in the fermionic loops. 

The elements studied here are general, defined by Fujikawa's method, and could be applied to a topological phase of gravity \cite{Witten:1988xi, Junqueira:2016hlu, Sadovski:2024uhg}, which was recently related to some puzzles concerning early universe cosmology \cite{Agrawal:2020xek}. The method could be applied to theories with other gauge symmetries. In Seiberg-Witten theory \cite{Seiberg:1994rs}, the monopole condensation occurs due to a symmetry breaking of the type $SU(2) \rightarrow U(1)$, which cannot be applied in this model, since the $SU(2)$ possess only one generator in Cartan subalgebra. We also pointed out that TQFTs have connections with conformal field theories. In particular, there is a quantum correspondence between the twisted $\mathcal{N}=2$ SYM and the Baulieu-Singer conformal gauge theory\cite{Junqueira:2021rwc}. The correlation between TQFTs and the AdS/CFT correspondence was studied in \cite{Witten:1998wy, BenettiGenolini:2017zmu}, which indicates that topological phases in field theory could reveal aspects in high-energy physics, from the application of effective models. Higgs bosons can appear in supersymmetric models, see for instance \cite{Gunion:1984yn, Gunion:1986nh}. Here, due to the mixture between bosonic and fermonic fields present in terms of topological Witten's action, the mass generation is transferred to the fermionic mass poles.    In general lines, such SSB provides useful tools to generate an energy scale that makes possible the investigation of quantum effects of broken topological phases in non-Abelian field theories, through the release of local degrees of freedom. In the present case, we have demonstrated that the introduction of a topological phase could provide a mechanism of mass generation for fermions and gauge bosons, with their masses defined by the same energy scale. It remains to demonstrate the renormalizability and unitarity of the theory after SSB. We must consider that, before the break, the observables are global, given by exact numbers. Usually, the introduction of BRST-exact terms does not destroy the quantum stability of the model, depending on the gauge-fixing choice. These issues are under investigation.

\section*{Acknowledgments}

OCJ thanks the S\~ao Paulo Research Foundation (FAPESP) 
(Grants No. 2021/01089-1 and No. 2024/14390-0). This work also received support from the National Council for Scientific and Technological Development - CNPq and   Coordena\c c\~ao de Aperfei\c coamento de Pessoal de N\'ivel Superior - Brasil (CAPES) - Finance Code 001. OCJ also thanks G. Sadovski for the useful discussions.

\bibliographystyle{utphys2}
\bibliography{library}

\providecommand{\href}[2]{#2}\begingroup\raggedright\begin{thebibliography}{10}

\bibitem{Belavin:1975fg}
A.~A. Belavin, A.~M. Polyakov, A.~S. Schwartz, and Y.~S. Tyupkin, ``{Pseudoparticle solutions of the Yang-Mills equations}''. \href{http://dx.doi.org/10.1016/0370-2693(75)90163-X}{{\em Physics Letters B} {\bfseries 59} no.~1, (1975) 85--87}.

\bibitem{DONALDSON1990257}
S.~Donaldson, ``Polynomial invariants for smooth four-manifolds''. \href{http://dx.doi.org/https://doi.org/10.1016/0040-9383(90)90001-Z}{{\em Topology} {\bfseries 29} no.~3, (1990) 257 -- 315}. \url{http://www.sciencedirect.com/science/article/pii/004093839090001Z}.

\bibitem{Donaldson:1983wm}
S.~K. Donaldson, ``{An application of gauge theory to four-dimensional topology}''. \href{http://dx.doi.org/10.4310/jdg/1214437665}{{\em Journal of Differential Geometry} {\bfseries 18} no.~2, (1983) 279--315}.

\bibitem{donaldson1987}
S.~K. Donaldson, ``The orientation of yang-mills moduli spaces and 4-manifold topology''. \href{http://dx.doi.org/10.4310/jdg/1214441485}{{\em J. Differential Geom.} {\bfseries 26} no.~3, (1987) 397--428}. \url{https://doi.org/10.4310/jdg/1214441485}.

\bibitem{Atiyah:1987ri}
M.~Atiyah, ``{NEW INVARIANTS OF THREE-DIMENSIONAL AND FOUR-DIMENSIONAL MANIFOLDS}''. {\em Proc. Symp. Pure Math.} {\bfseries 48} (1988) 285--299.

\bibitem{Witten:1988ze}
E.~Witten, ``{Topological quantum field theory}''. \href{http://dx.doi.org/10.1007/BF01223371}{{\em Communications in Mathematical Physics} {\bfseries 117} no.~3, (9, 1988) 353--386}.

\bibitem{West:1990}
P.~West, \href{http://dx.doi.org/10.1142/1002}{{\em {Introduction to Supersymmetry and Supergravity}}}.
\newblock World Scientific, 5, 1990.

\bibitem{Baulieu:1988xs}
L.~Baulieu and I.~Singer, ``{Topological Yang-Mills symmetry}''. \href{http://dx.doi.org/10.1016/0920-5632(88)90366-0}{{\em Nuclear Physics B - Proceedings Supplements} {\bfseries 5} no.~2, (12, 1988) 12--19}.

\bibitem{Brooks:1988jm}
R.~Brooks, D.~Montano, and J.~Sonnenschein, ``{Gauge fixing and renormalization in topological quantum field theory}''. \href{http://dx.doi.org/10.1016/0370-2693(88)90458-3}{{\em Physics Letters B} {\bfseries 214} no.~1, (11, 1988) 91--97}.

\bibitem{Birmingham:1988ap}
D.~Birmingham, M.~Rakowski, G.~Thompson, I.~Centre, T.~Pto, T.~Physics, I.~Centre, T.~Pttvsk, P.~Vi, and P.~Jussieu, ``{Renormalization of topological field theory}''. \href{http://dx.doi.org/10.1016/0550-3213(90)90058-L}{{\em Nuclear Physics B} {\bfseries 329} no.~1, (1, 1990) 83--97}.

\bibitem{WerneckdeOliveira:1993pa}
M.~Werneck~de Oliveira, ``{Algebraic renormalization of the topological Yang-Mills field theory}''. \href{http://dx.doi.org/10.1016/0370-2693(93)90231-6}{{\em Phys. Lett. B} {\bfseries 307} (1993) 347--352}.

\bibitem{Brandhuber:1994uf}
A.~Brandhuber, O.~Moritsch, M.~de~Oliveira, O.~Piguet, and M.~Schweda, ``{A renormalized supersymmetry in the topological Yang-Mills field theory}''. \href{http://dx.doi.org/10.1016/0550-3213(94)90102-3}{{\em Nuclear Physics B} {\bfseries 431} no.~1-2, (12, 1994) 173--190}, \href{http://arxiv.org/abs/hep-th/9407105}{{\ttfamily hep-th/9407105}}.

\bibitem{Blasi:2000qw}
A.~Blasi, V.~Lemes, N.~Maggiore, S.~Sorella, A.~Tanzini, O.~Ventura, and L.~Vilar, ``{Perturbative beta function of N=2 superYang-Mills theories}''. \href{http://dx.doi.org/10.1088/1126-6708/2000/05/039}{{\em JHEP} {\bfseries 05} (2000) 039}.

\bibitem{Junqueira:2017zea}
O.~C. Junqueira, A.~D. Pereira, G.~Sadovski, R.~F. Sobreiro, and A.~A. Tomaz, ``{Topological Yang-Mills theories in self-dual and anti-self-dual Landau gauges revisited}''. \href{http://dx.doi.org/10.1103/PhysRevD.96.085008}{{\em Physical Review D} {\bfseries 96} no.~8, (10, 2017) 085008}, \href{http://arxiv.org/abs/1707.06666}{{\ttfamily 1707.06666}}.

\bibitem{Junqueira:2018xgl}
O.~C. Junqueira, A.~D. Pereira, G.~Sadovski, R.~F. Sobreiro, and A.~A. Tomaz, ``{Absence of radiative corrections in four-dimensional topological Yang-Mills theories}''. \href{http://dx.doi.org/10.1103/PhysRevD.98.021701}{{\em Physical Review D} {\bfseries 98} no.~2, (7, 2018) 21701}, \href{http://arxiv.org/abs/1805.01850}{{\ttfamily 1805.01850}}.

\bibitem{Junqueira:2018zxr}
O.~C. Junqueira, A.~D. Pereira, G.~Sadovski, R.~F. Sobreiro, and A.~A. Tomaz, ``{More about the renormalization properties of topological Yang-Mills theories}''. \href{http://dx.doi.org/10.1103/PhysRevD.98.105017}{{\em Physical Review D} {\bfseries 98} no.~10, (11, 2018) 105017}, \href{http://arxiv.org/abs/1807.01517}{{\ttfamily 1807.01517}}.

\bibitem{Dudal:2019bjh}
D.~Dudal, C.~Felix, O.~Junqueira, D.~Montes, A.~Pereira, G.~Sadovski, R.~Sobreiro, and A.~Tomaz, ``{Infinitesimal Gribov copies in gauge-fixed topological Yang-Mills theories}''. \href{http://dx.doi.org/10.1016/j.physletb.2020.135531}{{\em Phys. Lett. B} {\bfseries 807} (2020) 135531}.

\bibitem{Junqueira:2021rwc}
O.~C. Junqueira and R.~F. Sobreiro, ``{Correspondence between the twisted $N=2$ super-Yang-Mills and conformal Baulieu-Singer theories}''. \href{http://dx.doi.org/10.1103/PhysRevD.103.085008}{{\em Phys. Rev. D} {\bfseries 103} no.~8, (2021) 085008}.

\bibitem{Witten:1988xi}
E.~Witten, ``{Topological Gravity}''. \href{http://dx.doi.org/10.1016/0370-2693(88)90704-6}{{\em Physics Letters B} {\bfseries 206} no.~4, (6, 1988) 601--606}.

\bibitem{vanBaal:1989aw}
P.~Van~Baal, ``{An introduction to Topological Yang-Mills Theory}''. {\em Acta Physica Polonica} {\bfseries B21} no.~2, (1990) 73.

\bibitem{Fujikawa:1982ss}
K.~Fujikawa, ``{Dynamical Stability of the {BRS} Supersymmetry and the Gribov Problem}''. \href{http://dx.doi.org/10.1016/0550-3213(83)90102-5}{{\em Nucl. Phys. B} {\bfseries 223} (1983) 218--234}.

\bibitem{Witten:1988hf}
E.~Witten, ``{Quantum field theory and the Jones polynomial}''. \href{http://dx.doi.org/10.1007/BF01217730}{{\em Communications in Mathematical Physics} {\bfseries 121} no.~3, (9, 1989) 351--399}.

\bibitem{Witten:1998wy}
E.~Witten, ``{AdS / CFT correspondence and topological field theory}''. \href{http://dx.doi.org/10.1088/1126-6708/1998/12/012}{{\em JHEP} {\bfseries 12} (1998) 012}.

\bibitem{BenettiGenolini:2017zmu}
P.~Benetti~Genolini, P.~Richmond, and J.~Sparks, ``{Topological AdS/CFT}''. \href{http://dx.doi.org/10.1007/JHEP12(2017)039}{{\em JHEP} {\bfseries 12} (2017) 039}.

\bibitem{PhysRevLett.88.031601}
J.~Polchinski and M.~J. Strassler, ``Hard scattering and gauge/string duality''. \href{http://dx.doi.org/10.1103/PhysRevLett.88.031601}{{\em Phys. Rev. Lett.} {\bfseries 88} (Jan, 2002) 031601}. \url{https://link.aps.org/doi/10.1103/PhysRevLett.88.031601}.

\bibitem{Boschi-Filho:2002wdj}
H.~Boschi-Filho and N.~R.~F. Braga, ``{QCD / string holographic mapping and glueball mass spectrum}''. \href{http://dx.doi.org/10.1140/epjc/s2003-01526-4}{{\em Eur. Phys. J. C} {\bfseries 32} (2004) 529--533}.

\bibitem{Boschi-Filho:2002xih}
H.~Boschi-Filho and N.~R.~F. Braga, ``{Gauge / string duality and scalar glueball mass ratios}''. \href{http://dx.doi.org/10.1088/1126-6708/2003/05/009}{{\em JHEP} {\bfseries 05} (2003) 009}.

\bibitem{PhysRevD.74.015005}
A.~Karch, E.~Katz, D.~T. Son, and M.~A. Stephanov, ``Linear confinement and ads/qcd''. \href{http://dx.doi.org/10.1103/PhysRevD.74.015005}{{\em Phys. Rev. D} {\bfseries 74} (Jul, 2006) 015005}. \url{https://link.aps.org/doi/10.1103/PhysRevD.74.015005}.

\bibitem{Herzog:2006ra}
C.~P. Herzog, ``{A Holographic Prediction of the Deconfinement Temperature}''. \href{http://dx.doi.org/10.1103/PhysRevLett.98.091601}{{\em Phys. Rev. Lett.} {\bfseries 98} (2007) 091601}.

\bibitem{PhysRevD.77.046002}
C.~A.~B. Bayona, H.~Boschi-Filho, N.~R.~F. Braga, and L.~A.~P. Zayas, ``On a holographic model for confinement/deconfinement''. \href{http://dx.doi.org/10.1103/PhysRevD.77.046002}{{\em Phys. Rev. D} {\bfseries 77} (Feb, 2008) 046002}. \url{https://link.aps.org/doi/10.1103/PhysRevD.77.046002}.

\bibitem{Braga:2022yfe}
N.~R.~F. Braga, L.~F. Faulhaber, and O.~C. Junqueira, ``{Confinement-deconfinement temperature for a rotating quark-gluon plasma}''. \href{http://dx.doi.org/10.1103/PhysRevD.105.106003}{{\em Phys. Rev. D} {\bfseries 105} no.~10, (2022) 106003}.

\bibitem{Weis:1997kj}
M.~Weis, ``{Topological Aspects of Quantum Gravity}''. \href{http://arxiv.org/abs/hep-th/9806179}{{\ttfamily hep-th/9806179}}.

\bibitem{Delduc:1996yh}
F.~Delduc, N.~Maggiore, O.~Piguet, and S.~Wolf, ``{Note on constrained cohomology}''. \href{http://dx.doi.org/10.1016/0370-2693(96)00879-9}{{\em Phys. Lett. B} {\bfseries 385} (1996) 132--138}.

\bibitem{Boldo:2003jq}
I.~S. Boldo, C.~P. Constantinidis, O.~Piguet, M.~Lefranc, J.~L. Boldo, C.~P. Constantinidis, F.~Gieres, M.~Lefrancois, and O.~Piguet, ``{Observables in Topological Yang-Mills Theories}''. \href{http://dx.doi.org/10.1142/S0217751X0401777X}{{\em International Journal of Modern Physics A} {\bfseries 19} no.~17n18, (3, 2003) 2971--3004}, \href{http://arxiv.org/abs/hep-th/0303053}{{\ttfamily hep-th/0303053}}.

\bibitem{JMKosterlitz_1972}
J.~M. Kosterlitz and D.~J. Thouless, ``Long range order and metastability in two dimensional solids and superfluids. (application of dislocation theory)''. \href{http://dx.doi.org/10.1088/0022-3719/5/11/002}{{\em Journal of Physics C: Solid State Physics} {\bfseries 5} no.~11, (Jun, 1972) L124}. \url{https://dx.doi.org/10.1088/0022-3719/5/11/002}.

\bibitem{Kosterlitz:1973xp}
J.~M. Kosterlitz and D.~J. Thouless, ``{Ordering, metastability and phase transitions in two-dimensional systems}''. \href{http://dx.doi.org/10.1088/0022-3719/6/7/010}{{\em J. Phys. C} {\bfseries 6} (1973) 1181--1203}.

\bibitem{Thouless:1982zz}
D.~J. Thouless, M.~Kohmoto, M.~P. Nightingale, and M.~den Nijs, ``{Quantized Hall Conductance in a Two-Dimensional Periodic Potential}''. \href{http://dx.doi.org/10.1103/PhysRevLett.49.405}{{\em Phys. Rev. Lett.} {\bfseries 49} (1982) 405--408}.

\bibitem{Haldane:1983ru}
F.~D.~M. Haldane, ``{Nonlinear field theory of large spin Heisenberg antiferromagnets. Semiclassically quantized solitons of the one-dimensional easy Axis Neel state}''. \href{http://dx.doi.org/10.1103/PhysRevLett.50.1153}{{\em Phys. Rev. Lett.} {\bfseries 50} (1983) 1153--1156}.

\bibitem{Wess:1992cp}
J.~Wess and J.~Bagger, {\em {Supersymmetry and supergravity}}.
\newblock Princeton University Press, Princeton, NJ, USA, 1992.

\bibitem{Maggiore:1994dw}
N.~Maggiore, ``{Algebraic renormalization of N=2 superYang-Mills theories coupled to matter}''. \href{http://dx.doi.org/10.1142/S0217751X95001789}{{\em Int. J. Mod. Phys. A} {\bfseries 10} (1995) 3781--3802}.

\bibitem{Blasi:1989ka}
A.~Blasi and R.~Collina, ``{Basic Cohomology of Topological Quantum Field Theories}''. \href{http://dx.doi.org/10.1016/0370-2693(89)90336-5}{{\em Phys. Lett. B} {\bfseries 222} (1989) 419--424}.

\bibitem{Higgs:1964pj}
P.~W. Higgs, ``{Broken Symmetries and the Masses of Gauge Bosons}''. \href{http://dx.doi.org/10.1103/PhysRevLett.13.508}{{\em Physical Review Letters} {\bfseries 13} no.~16, (10, 1964) 508--509}.

\bibitem{Dixon:1991wi}
J.~A. Dixon, ``{Calculation of BRS cohomology with spectral sequences}''. \href{http://dx.doi.org/10.1007/BF02101877}{{\em Commun. Math. Phys.} {\bfseries 139} (1991) 495--526}.

\bibitem{Piguet:1995er}
O.~Piguet and S.~P. Sorella, \href{http://dx.doi.org/10.1007/978-3-540-49192-7}{{\em {Algebraic Renormalization}}}, vol.~28 of {\em Lecture Notes in Physics Monographs}.
\newblock Springer Berlin Heidelberg, Berlin, Heidelberg, 1995.

\bibitem{Vandersickel:2011zc}
N.~Vandersickel, {\em {A study of the Gribov-Zwanziger action: from propagators to glueballs}}.
\newblock PhD thesis, Ghent University, 4, 2011.
\newblock \href{http://arxiv.org/abs/1104.1315}{{\ttfamily 1104.1315}}.

\bibitem{Junqueira:2016hlu}
O.~C. Junqueira, A.~D. Pereira, G.~Sadovski, T.~R.~S. Santos, R.~F. Sobreiro, and A.~A. Tomaz, ``{Equivalence between the Lovelock–Cartan action and a constrained gauge theory}''. \href{http://dx.doi.org/10.1140/epjc/s10052-017-4820-y}{{\em The European Physical Journal C} {\bfseries 77} no.~4, (4, 2017) 249}, \href{http://arxiv.org/abs/1612.05590}{{\ttfamily 1612.05590}}.

\bibitem{Sadovski:2024uhg}
G.~Sadovski and R.~F. Sobreiro, ``{Topological symmetry-restored phase of gravity}''. \href{http://dx.doi.org/10.1140/epjc/s10052-025-14274-y}{{\em Eur. Phys. J. C} {\bfseries 85} no.~6, (2025) 710}.

\bibitem{Agrawal:2020xek}
P.~Agrawal, S.~Gukov, G.~Obied, and C.~Vafa, ``{Topological Gravity as the Early Phase of Our Universe}''.

\bibitem{Seiberg:1994rs}
N.~Seiberg and E.~Witten, ``{Electric-magnetic duality, monopole condensation, and confinement in N=2 supersymmetric Yang-Mills theory}''. \href{http://dx.doi.org/10.1016/0550-3213(94)90124-4}{{\em Nuclear Physics B} {\bfseries 426} no.~1, (9, 1994) 19--52}, \href{http://arxiv.org/abs/hep-th/9407087}{{\ttfamily hep-th/9407087}}.

\bibitem{Gunion:1984yn}
J.~F. Gunion and H.~E. Haber, ``{Higgs Bosons in Supersymmetric Models. 1.}''. \href{http://dx.doi.org/10.1016/0550-3213(86)90340-8}{{\em Nucl. Phys. B} {\bfseries 272} (1986) 1}. [Erratum: Nucl.Phys.B 402, 567--569 (1993)].

\bibitem{Gunion:1986nh}
J.~F. Gunion and H.~E. Haber, ``{Higgs Bosons in Supersymmetric Models. 2. Implications for Phenomenology}''. \href{http://dx.doi.org/10.1016/0550-3213(86)90050-7}{{\em Nucl. Phys. B} {\bfseries 278} (1986) 449}. [Erratum: Nucl.Phys.B 402, 569--569 (1993)].

\end{thebibliography}\endgroup

\end{document}